\documentclass[11pt,a4paper]{article}

\usepackage[T1]{fontenc}
\usepackage[utf8]{inputenc}   %
\usepackage{lmodern}          %
\usepackage{microtype}        %

\usepackage{amsmath}
\usepackage{amssymb}
\usepackage{bm}

\usepackage{siunitx}
\usepackage{graphicx}
\usepackage{caption}
\usepackage{subcaption}
\usepackage{float}

\usepackage{booktabs}
\usepackage{multirow}

\usepackage[numbers,sort&compress]{natbib}

\usepackage[hidelinks]{hyperref}

\usepackage{geometry}
\title{Thermoelastic harvesting outperforming thermoelectric generators below 100\,°C}

\author{%
Bruno Neumann$^{1,2}$,
Andreas Henschke$^{1}$,
Morik Nikolic$^{1,3}$,
Sebastian Fähler$^{1}$\thanks{Corresponding author: s.faehler@hzdr.de}\\[0.75em]
\small $^{1}$Institute of Ion Beam Physics and Materials Research,\\
\small \phantom{$^{1}$}Helmholtz-Zentrum Dresden--Rossendorf, Dresden, Germany\\
\small $^{2}$Institute of Process Engineering and Environmental Technology,\\
\small \phantom{$^{2}$}TUD Dresden University of Technology, 01069 Dresden, Germany\\
\small $^{3}$Hochschule für Technik und Wirtschaft Dresden, Dresden, Germany
}

\date{}

\begin{document}
\maketitle

\begin{abstract}
Low-grade waste heat below 100\,°C is one of the largest untapped opportunities
in solid-state energy conversion. Three ferroic routes are candidates for recovering this resource:
thermomagnetic, pyroelectric, and thermoelastic harvesters.
The last has remained the most unexplored, despite decades of progress on the underlying NiTi shape-memory alloy wires. Three system-design changes close this gap:
a protagonist--antagonist architecture that recovers the energy for prestraining,
a continuously tunable prestrain mechanism that sets the force--strain balance, and transversal water flow that decouples cycle frequency from wire length. The resulting harvester delivers a directly measured power density of 366\,mW/cm³ with respect to the active material, about 1.7 times the next-best thermoelastic device,
ahead of every reported thermomagnetic and pyroelectric generator, and outperforming the best thermoelectric generators in this temperature range also with respect to power per material cost. The system maps the parameter space directly through force and displacement measurements, without using material-property estimates, giving a quantitative picture of how the alloy responds while the device is doing work.
\end{abstract}

\section*{Introduction}
In a heating world,\cite{Zhang2026}
action is needed on two fronts:
more efficient cooling with a low carbon footprint,
and the recovery of the low-grade waste heat
that human activity today dissipates at unprecedented scale.\cite{Forman2016}
Ferroic materials, comprising ferromagnetic, ferroelectric and ferroelastic materials
and their combinations, can address both
as conjugate forms of solid-state energy conversion.
Each material class exhibits a phase transition that can be exploited in either thermodynamic direction.
Run in one direction, the transition releases heat under an applied field, enabling cooling:
the magnetocaloric,\cite{Liu2012} electrocaloric\cite{Li2023} and elastocaloric\cite{Tusek2016} effects,
together with multicaloric variants that combine several order parameters.\cite{faehler2012}
Run in the other direction, a temperature change drives the transition
and the conjugate field can be converted to electric or mechanical work,
giving the corresponding harvesters:
thermomagnetic,\cite{Bahl2024,Waske2019,Gueltig2017}
pyroelectric\cite{Lheritier2022}
and thermoelastic.\cite{LE1978,Sato2008,Kumar2019}

Cooling is currently the more mature of the two routes:
the first commercial magnetocaloric heat pumps have entered the market\cite{Liang2026}
and electrocaloric, elastocaloric and multicaloric demonstrators are approaching the same milestone,
offering solid-state alternatives to the greenhouse-gas refrigerants of vapor compression.
The thermodynamic counterpart, ferroic harvesting,
has lagged far behind, despite a much larger potential impact:
cooling accounts for about 20\,\% of global primary energy use,\cite{IEA2018}
whereas about two thirds of all primary energy
is ultimately dissipated as heat,\cite{Firth2019,Forman2016,Papapetrou2018}
most of it below 100\,°C\cite{Firth2019,Forman2016},
a temperature range in which no economically viable conversion technology is available.
The volume of low-grade waste heat is projected to grow further,
driven in particular by the thermal output of data centres serving AI workloads.\cite{Hao2025,Yuan2025}

The most mature non-ferroic solid-state option,
thermoelectric generators (TEG),\cite{Jaziri2020,Luo2014,Champier2017,Kishore2018,SaufiSulaiman2019}
reach only a small fraction of Carnot efficiency at low temperature difference $\Delta T$
and depend on scarce or toxic elements such as Bi and Te.
Among mechanical alternatives, Stirling cycles\cite{Wang2016}
introduce rotating machinery
that erodes the lifetime advantage of solid-state conversion.
Organic Rankine cycles (ORC) share this drawback,
and although ultra-low-temperature working fluids are under active study,
demonstrated devices below 100\,°C are still absent.\cite{Cao2023b}
Among solid-state ferroic harvesters,
thermomagnetic and pyroelectric generators
have been demonstrated below 100\,°C
but still trail TEG on power density.
The third ferroic route, \textbf{thermoelastic harvesting} (TEH) using shape-memory alloy (SMA) wires,
has remained the weakest of the three
despite being proposed almost fifty years ago.\cite{LE1978}
The alloys it relies on, NiTi in particular,\cite{Buehler1963}
have meanwhile matured through decades of medical-device development.\cite{Chluba2015,Hou2019}
When changing its temperature, SMA
switch between two markedly different mechanical regimes:
a stiff austenite at high temperature
and a martensite easily deformed by moving twin boundaries at low temperature.\cite{Otsuka1999,Huber1997}
The performance of a TEH cannot be separated from the SMA inside it.
Our focus here is the interaction between device and material under harvesting conditions,
a thread we develop through the results that follow.

Three specific system-design choices have kept TEH below the rest of the field.
First, most demonstrators follow Johnson's 1976 patent\cite{Johnson1976}
of an endless wire looped between a hot and a cold pulley,\cite{Sato2008,Kumar2019}
a layout in which the wire rotates continuously between two baths
and passes through only one hot--cold interface per revolution,
so that at any instant only the narrow segment currently crossing this interface is actively transforming
and the rest of the active material sits idle in either bath.
Second, the mechanical work an SMA wire delivers per cycle
is the product of two coupled quantities, force and strain,
whose balance is set by the prestrain at which the wire is operated.
Prior designs fix the prestrain through their mechanical geometry,
so neither force nor strain can be optimized in operation,
and previous works have addressed only one of the two.
Third, prior designs do not scale up:
axial fluid flow along the wire couples wire length to heat-exchange time,
so increasing the active volume, the natural route to higher power,
slows the cycle, and long wires and fast cycles
cannot be achieved together.
None of these limits is fundamental: each follows from a single engineering choice.

We address all three.
First, we build a protagonist--antagonist generator following Sakuma et al.,\cite{SAKUMA1998}
in which two batches of SMA wires alternate discretely between hot and cold,
so that the full active volume transforms each half-cycle
rather than only the segment currently at the hot--cold interface.
Second, we integrate our recently patented prestrain mechanism,\cite{Neumann2020}
which makes the prestrain a continuously tunable control parameter.
Third, we replace axial with transversal fluid flow,\cite{Neumann2024,Neumann2025a}
decoupling heat exchange from wire length
so that long wires and fast cycles, both required for high power,
can be combined.
We characterize the resulting device
by first sketching the underlying thermodynamic cycle,
then mapping the actuation parameter space without load,
then optimizing output power under load,
and finally benchmarking against all prior ferroic harvesters and against TEG below 100\,°C.
The TEH reaches a power density of 366\,mW/cm³
and a material-cost-normalized power of 4.9\,W/€,
outperforming every previously reported TEH, thermomagnetic generator, pyroelectric generator,
and thermoelectric generator in this temperature range.
This work focuses on the system-design problem of TEH,
namely device architecture and operating point,
and treats the complementary material-design problem in the outlook.

\begin{figure}[H]
  \centering
  \makebox[\textwidth][c]{%
    \includegraphics[width=1\textwidth]{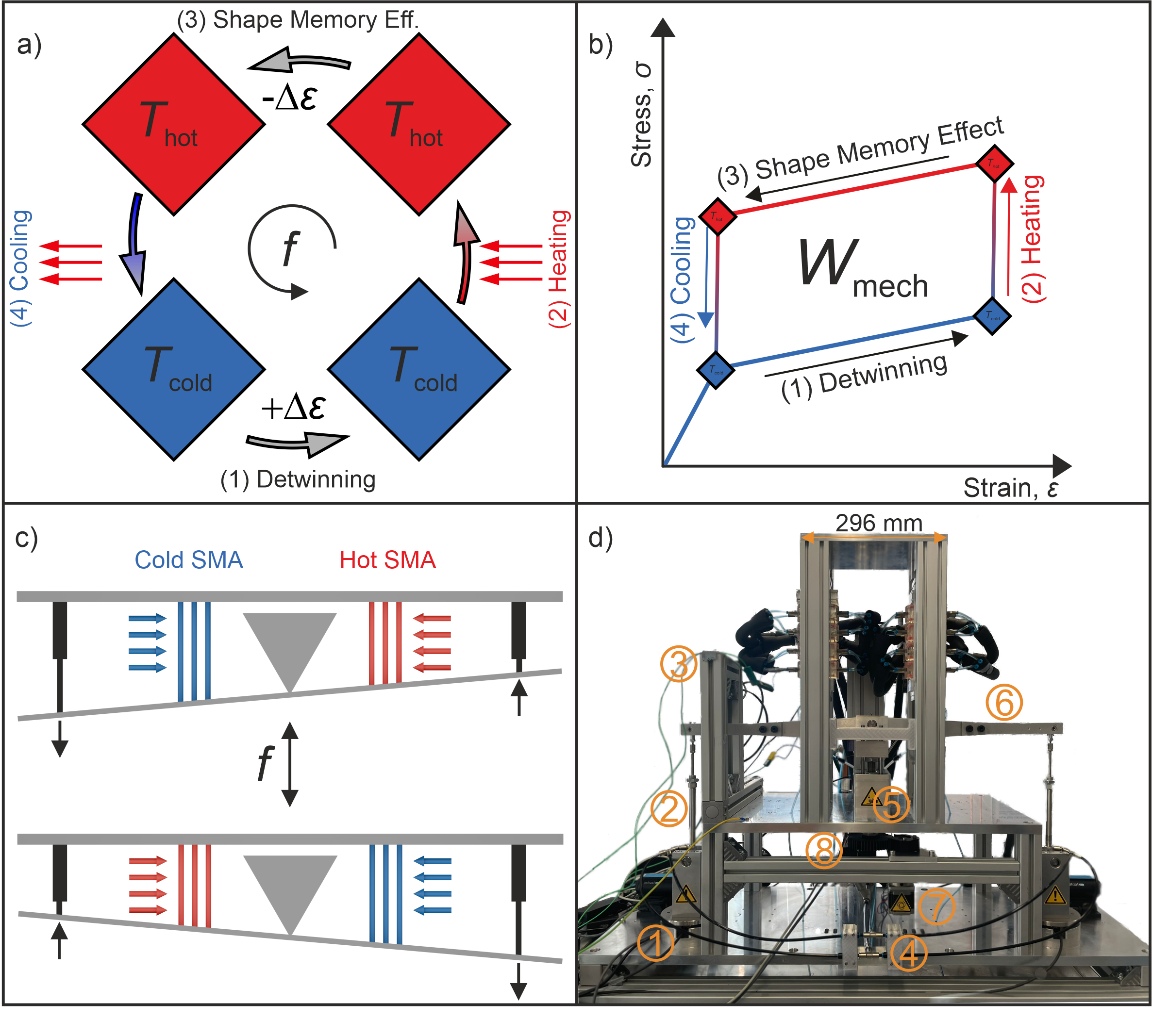}
  }
  \caption{\textbf{From the thermoelastic harvesting cycle to a powerful thermoelastic harvester of low-grade waste heat using shape memory alloys (SMA)}.  a) Four stages of a thermoelastic cycle, which is driven by waste heat with a temperature $T_{\text{hot}}$ with respect to ambient $T_{\text{cold}}$ to actuate SMA. b) A schematic temperature dependent stress-strain diagram illustrates, how these four stages create mechanical work $W_{\text{mech}}$. c) Sketch of a protagonist-antagonist thermoelastic harvester (TEH), where two sets of SMA wires are connected to a seesaw at the bottom. By alternating their temperature between $T_{\text{hot}}$ and $T_{\text{cold}}$ at the cycle frequency \textit{f}, the seesaw tilts between left and right, as shown in top and bottom sketches, respectively. Hydraulic cylinders extract mechanical work from this reciprocating heat engine. The central bearing can be moved up and down to vary
  the prestrain on the SMA wires. d) Photo of the TEH, with the major instrumentation labeled. 1: Force sensor, 2: Hydraulic cylinder, 3: Displacement sensor, 4: Oil circuit, 5: Central bearing, 6: Seesaw, 7: Stepper motor for prestrain, 8: Throttle valve. A video showing the reciprocating movement of the seesaw is provided as Movie~S1 in the supplementary information.}
  \label{fig:Setup}
\end{figure}

\section*{Results}
\subsection*{From the harvesting cycle to a powerful thermoelastic device}
\label{sec:Setup}

A ferroic harvester runs a thermodynamic cycle in which a phase transition is
driven by alternating contact with a hot source at $T_\mathrm{hot}$ and a cold sink at $T_\mathrm{cold}$.
The thermal input to the cycle is
$\dot{Q}_\mathrm{in} = \dot{V}\,\rho\,c_p\,\Delta T$,
with $\Delta T = T_\mathrm{hot} - T_\mathrm{cold}$,
$c_p$ the heat capacity, $\rho$ the density, and $\dot{V}$ the volume flow of the heat-transfer fluid.
Across each cycle the phase transition releases a conjugate ferroic field that the device converts into useful work: for thermomagnetic generators the field is magnetic, for pyroelectric generators it is electric, and for thermoelastic harvesters it is mechanical stress with strain as the conjugate property.

For a thermoelastic harvester, the cycle traces a closed loop on the stress-strain plane of the SMA.
Fig.~\ref{fig:Setup}a sketches the four stages of an idealised cycle,
and Fig.~\ref{fig:Setup}b shows the corresponding loop on the stress-strain diagram.
In stage 1 the wire is strained in the cold martensitic state at low stress,
exploiting the easy mobility of twin boundaries.\cite{Otsuka1999}
In stage 2, heating to the stiff austenite state raises the stress at almost constant strain.
In stage 3 the wire contracts under load, doing work against this load.
In stage 4 the wire is cooled, the stress relaxes, and the cycle closes.
The enclosed area is the mechanical work $W_\mathrm{mech}$ delivered per cycle,
so the output power is $P = W_\mathrm{mech}\,f$, with $f$ being the cycle frequency.
The loop in Fig.~\ref{fig:Setup}b looks deceptively like an isothermal pseudoelastic stress-strain curve,
but it is driven by alternating temperature rather than by strain alone,
as indicated by the blue and red colors.
$W_\mathrm{mech}$ therefore depends not on the SMA alone
but on where the operating cycle is placed
within the full stress-strain response of the SMA.
The prestrain, the temperature span $\Delta T$, and the applied load
each shift the cycle to a different region of this response
and change its enclosed area.
This is the first point at which the device-material coupling becomes visible.
A detailed finite-element treatment of the coupled thermo-elastic problem is given in our companion work.\cite{Neumann2025}

The three that follow are system-design choices that fix the device architecture and the operating point within which the SMA is then asked to work.
The complementary material-design question is taken up in the outlook.
We implement this cycle as the protagonist-antagonist generator shown in Fig.~\ref{fig:Setup}c.
Two sets of SMA wires are heated and cooled alternately and connected by a central seesaw,
so the work required to re-strain the cold side
is supplied mechanically by the hot side rather than from outside the system.
The same principle keeps the magnetic or electric field within the device
in recent thermomagnetic\cite{Waske2019} and pyroelectric\cite{Lheritier2022} harvesters.
Hydraulic cylinders at the outer edges of the seesaw extract the mechanical work.
To exchange heat between fluid and wires we use transversal water flow,\cite{Neumann2024,Neumann2025a}
which removes the axial heat-transfer bottleneck and decouples cycle frequency from wire length.
The axial bottleneck is the shortcoming of both the canonical Johnson-patent layout and the first implementation of the protagonist--antagonist concept by Sakuma et al.\cite{SAKUMA1998}
The central bearing of the seesaw can be raised or lowered
to set the prestrain $\epsilon_0$ on the wires.\cite{Neumann2020}
Because protagonist and antagonist are mechanically coupled through the seesaw,
the prestrain shifts the operating cycle of both wires together
along the SMA stress-strain response,
controlling the stress and the strain amplitude that each wire develops per cycle.
Identifying the prestrain that maximizes the work output $W_\mathrm{mech}$
is one of the central experimental tasks of this paper.
The prestrain is sketched in the inset of Fig.~\ref{fig:Strainamplitude}a,
which also introduces the strain amplitude $\Delta\epsilon$ used to describe the seesaw motion in the following.
The total prestrain $\epsilon_0$ is set by moving the central bearing,
which acts on both wires simultaneously.
Its distribution between the two wires becomes well defined during cycling,
when the first martensite-to-austenite transformation on the hot side extends the opposite wires to their maximum.

The device is realized with four commercially available NiTi wires, two on each side,
of 1\,mm diameter and 165\,mm active length.
Water serves as the heat-transfer fluid, since low-grade waste heat is most often carried by water.\cite{Forman2016}
The hydraulic system, fluid management, and the position and force sensors used to measure the mechanical work
are labeled in Fig.~\ref{fig:Setup}c and described in detail in the methods section.
Throughout this paper we convert measured forces and displacements into engineering stress and strain on the SMA,
using the known wire and seesaw geometry,
so that device results can be compared directly with the material stress-strain curve.
Two features of this implementation matter for what follows.
First, every process parameter that determines TEH performance,
namely $T_\mathrm{hot}$, $T_\mathrm{cold}$, $f$, $\epsilon_0$ and the load,
is directly controllable.
Second, both the thermal input $\dot{Q}_\mathrm{in}$ and the mechanical output $\bar{P}$ are measured directly,
without using material-property estimates.
The upper bound on conversion efficiency is the Carnot limit $1 - T_\mathrm{cold}/T_\mathrm{hot}$,
and we report device efficiency throughout as
$\eta_\mathrm{sys}/\eta_\mathrm{Carnot} = (\bar{P}/\dot{Q}_\mathrm{in})/(1 - T_\mathrm{cold}/T_\mathrm{hot})$.
We use this controllability in the sections that follow
to identify, step by step, the design rules for a powerful TEH
and the device-material couplings they reveal.
Conversion of the extracted mechanical work to electricity can be solved by standard electromagnetic generators
with efficiencies of 90\,\% and more,\cite{Chapman2012} and is therefore not part of this work.

\subsection*{Maximizing strain amplitude without load}
\label{sec:opencircuit}

As described in the previous section, strain is the decisive parameter for a TEH. Accordingly, we analyze
the influence of the key process parameters on the strain amplitude $\Delta\epsilon$ and identify the optimum, feasible process parameters. As starting values we take the limits of our device and proceed sequentially: we take the optimum value from each step and keep it constant for the next. To keep the text readable, we give all values measured only within the figures. These experiments are performed without extracting work, achieved by removing the hydraulic cylinders.

\begin{figure}[H]
  \centering
  \makebox[\textwidth][c]{%
    \includegraphics[width=1.25\textwidth]{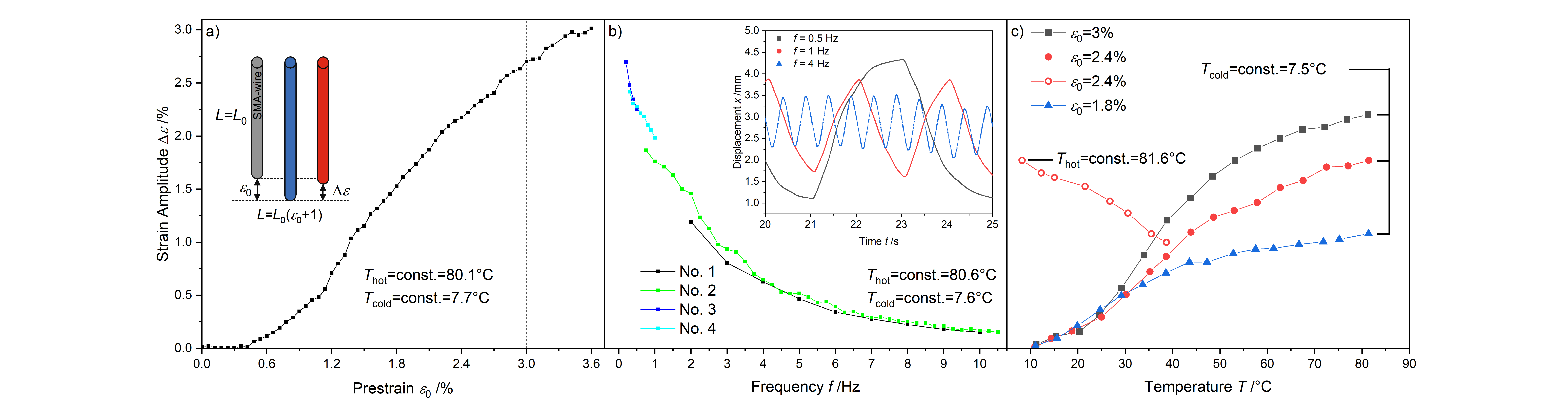}
  }  \caption{\textbf{Maximizing the strain amplitude $\Delta\epsilon$ of a TEH without load.} a) The prestrain quantifies how far the SMA wire is extended from its initial state. Cycling the temperature between hot and cold then yields the strain amplitude $\Delta\epsilon$. b) Influence of the cycling frequency $f$. These experiments were repeated with four different wires, marked by their number. The inset shows the underlying time dependence of the seesaw displacement at three different frequencies. c) Variation of $T_\mathrm{hot}$ and $T_\mathrm{cold}$, where the other temperature was held constant at its extreme value. The variation of $T_\mathrm{cold}$ was probed at $\epsilon_0 = 2.4$\,\%, and for $T_\mathrm{hot}$ at three different prestrain values. The specific values measured during each run are given within the figure panels throughout this paper.}
  \label{fig:Strainamplitude}
\end{figure}

As a first step, we examine the influence of the wire prestrain $\epsilon_0$, shown in Fig.~\ref{fig:Strainamplitude}a. Up to $\epsilon_0 = 0.25$\,\%, no strain amplitude is obtained, indicating that at this low prestrain the wire remains within its elastic regime, which does not differ much for austenite and martensite. A further increase produces a linear rise of $\Delta\epsilon$ that begins to level off around $\epsilon_0 = 3.6$\,\%. We did not increase $\epsilon_0$ further, since $5$\,\% is a typical strain at which the pseudoelastic plateau in NiTi reaches its limit.\cite{Otsuka1999} To mitigate fatigue, we reduced the prestrain to 3\,\% in the following experiments, where $\Delta\epsilon$ is 2.7\,\%, about half of the 5\,\% limit. This indicates that roughly half of the wire transforms between the austenite and martensite phases in each cycle, assuming transformation strain maps linearly onto transformed volume fraction.

As a second step, we investigate the influence of the cycle frequency $f$, which sets how fast we switch between hot and cold water. In the previous step, we kept $f$ as low as 0.25\,Hz to ensure an almost complete heat exchange between water and wire, but a high-power TEH requires a high cycle frequency. We observe a steep decrease of $\Delta\epsilon$ with $f$ (Fig.~\ref{fig:Strainamplitude}b). The frequency series was repeated on four separate wires, marked by their numbers in the figure, with a wire-to-wire spread of about 10\%. Although a small strain amplitude of 0.25\,\% is still measurable at 10\,Hz, we selected 0.5\,Hz for the following experiments, since this gives a much higher $\Delta\epsilon = 2.25$\,\%. To illustrate the origin of this strong frequency dependence, the inset depicts the temporal evolution of strain at three different frequencies. At 0.5\,Hz, $\Delta\epsilon$ approaches a stable value at the end of each half-cycle, showing that this time is sufficient for an almost complete heat exchange and the corresponding phase transformation. At 1\,Hz the plateau is no longer reached and the amplitude is reduced. At 4\,Hz, the strain amplitude decreases further and a sharp zig-zag curve is measured, which reveals that at this high frequency we switch between hot and cold water before the wire can reach the water temperature. Each switch stops the phase transformation in an incomplete state and reverses its direction. The incomplete heat exchange between fluid and wire leaves much of the heat unused, which is detrimental to the system efficiency.

As a third step, we vary the temperatures of the hot and cold water. Fig.~\ref{fig:Strainamplitude}c summarizes both series. Open symbols mark the series where $T_\mathrm{hot}$ was kept constant and $T_\mathrm{cold}$ varied, and vice versa for the closed symbols. In both series, $\Delta\epsilon$ decreases when the maximum temperature span is no longer used, indicating that this span is required to transform most of the material. The two series differ, however: $\Delta\epsilon$ decreases much faster when $T_\mathrm{cold}$ is increased, whereas an initial reduction of $T_\mathrm{hot}$ has little influence. To understand this asymmetry, we consider the wire's transformation temperature of $-10$\,°C, the mean of the four transformation temperatures given in the Methods Section. This transformation temperature lies not at the center of our temperature span but well below it. We deliberately selected wires with a low transformation temperature, because stress induces the martensite phase and therefore raises the transformation temperature under stress according to a Clausius--Clapeyron equation.\cite{Otsuka1999} In hindsight, wires with a transformation temperature 15\,K higher would have been a better choice, as it would have shifted $T_\mathrm{cold}$ closer to ambient as required for a harvesting system. The transformation temperatures of SMAs can be tuned through composition\cite{Frenzel2015} or heat treatment,\cite{KhalilAllafi2002} which will allow a TEH to be adapted to a broad range of waste-heat temperatures. The additional stress-induced shift of the transformation temperature is also visible in our experiments: varying the prestrain shifts the $\Delta\epsilon(T)$ curves to higher values, shown by the different colors of the closed symbols in Fig.~\ref{fig:Strainamplitude}c.

This load-free analysis identifies the prestrain, frequency, and temperature span
at which the strain amplitude is largest,
and probes how the SMA responds under device-controlled conditions.
It is the thermoelastic counterpart of measuring the open-circuit voltage
of a thermomagnetic or pyroelectric generator:
a necessary first step to understand the device, but not the regime in which it delivers power.
Most prior studies of thermoelastic actuation stop here.
In the next section we go further:
we apply a mechanical load and quantify the work and power delivered,
the regime in which a working TEH is actually judged.

\subsection*{From actuation to work}
A useful TEH must perform mechanical work, and in this section we identify the process conditions for optimum output power. As in the previous section, we proceed sequentially, using the parameters that maximized strain amplitude as the starting point.

\begin{figure}[H]
  \centering
  \makebox[\textwidth][c]{%
    \includegraphics[width=1\textwidth]{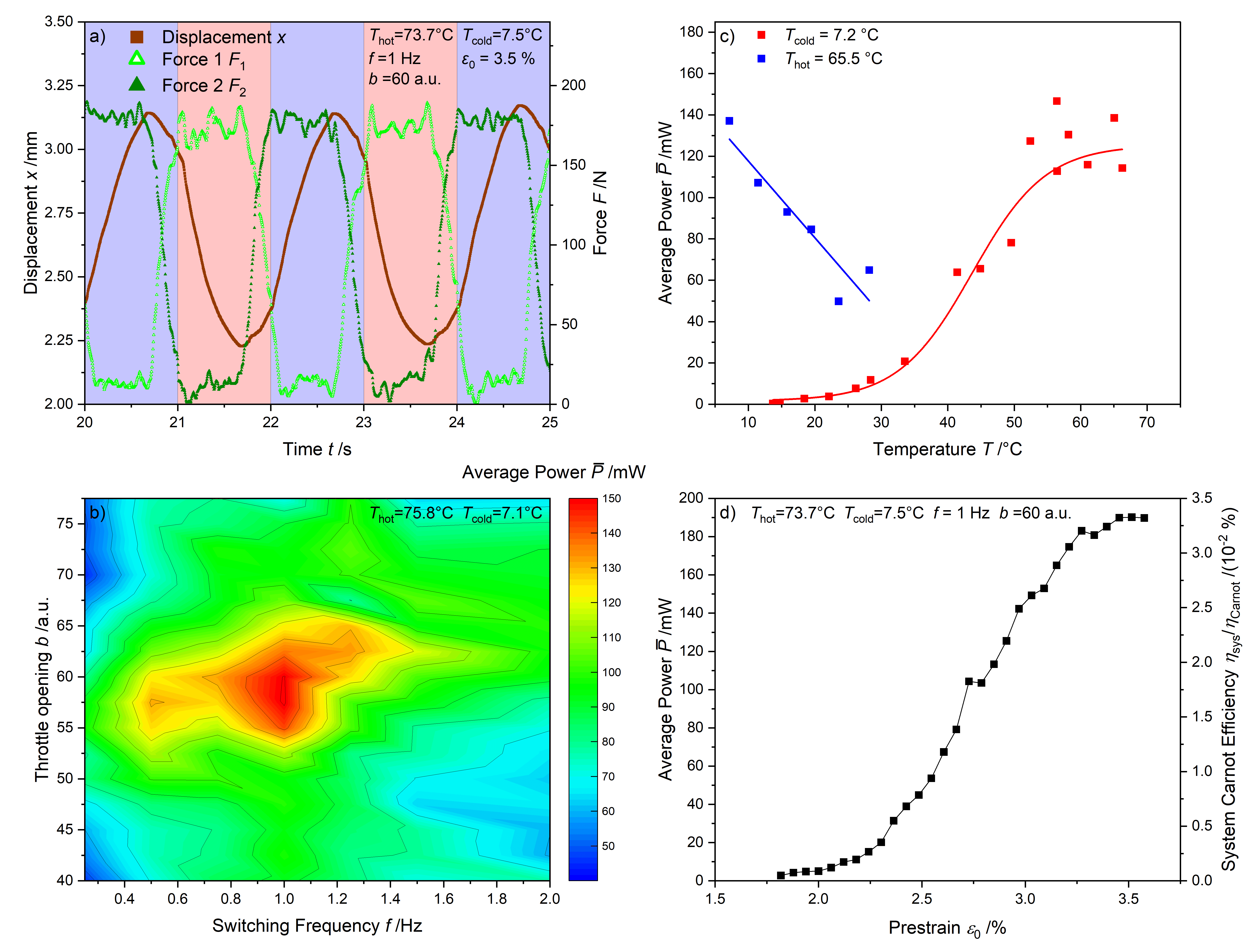}
  }

\caption{\textbf{Optimizing output power and efficiency of our TEH by varying process parameters.} a) Quantifying instantaneous output power by measuring displacement $x$ and forces $F_{i}$ at both sides of the protagonist-antagonist TEH. This representative measurement was performed at optimum average output power $\bar{P}$ of 190\,mW.  b) Identifying optimum frequency and throttle opening for maximum output power. c) Influence of hot and cold water temperature on output power. The corresponding temperature was kept constant. d) Prestraining to maximize output power. As additional axis on the right, the system efficiency in relation to Carnot efficiency is given.}
  	\label{fig:Outputpower}
\end{figure}

To extract work, we connect hydraulic cylinders to both sides of the seesaw; the cylinders pump oil through a variable throttle valve. We measure the displacement $x$ and the forces $F_1$ and $F_2$ on the two sides, as sketched in Fig.~\ref{fig:Setup}c. A representative measurement at optimum output power is shown in Fig.~\ref{fig:Outputpower}a. From these we obtain the instantaneous power $P_i = F_i \dot{x}$ on each side, and the output power $\bar{P}$ as the time average of their sum, with the averaging procedure detailed in supplementary Fig.~S1. This direct measurement of mechanical output, without using material-property estimates, is the central methodological feature of the section. The throttle valve serves only as a variable load whose opening we report in arbitrary units (a.u.), as it is not calibrated to absolute flow.

As a first step, we analyze the influence of the switching frequency $f$, since the previous load-free series did not yield an optimum value. The load applied by the throttle valve depends on the flow speed, so the throttle position and $f$ must be optimized together. Fig.~\ref{fig:Outputpower}b shows the resulting contour plot. We find a well-defined maximum output power at $f = 1$\,Hz and a throttle opening of 60\,a.u., which underlines how decisive both parameters are. The decrease at higher frequencies originates from the incomplete heat exchange between wire and water already established in the load-free series. The decrease at lower frequencies appears only here and not in the strain-amplitude series: its origin is again hot water leaving the system unused, but in this case because the long cycle time delivers more hot water than the wire can absorb.
These two loss mechanisms are absent at the optimum, as seen in the strain and force traces of Fig.~\ref{fig:Outputpower}a.
These traces correspond to the fully optimized operating point; the section walks through how it was reached. The force has a near-rectangular shape in each half-cycle, indicating that the SMA delivers force throughout the hot half cycle. The strain has a sinusoidal shape, as expected for a harvester running continuously. Force and strain in the same panel are also phase-shifted by approximately 90°, meaning force is in phase with velocity, the signature of a purely dissipative load at matched reactance.

As a second step, we analyze the influence of $T_\mathrm{hot}$ and $T_\mathrm{cold}$ on output power in the same way. The trends mirror those of the strain amplitude: reducing the temperature span also reduces the output power. The difference is, however, more pronounced. Increasing $T_\mathrm{cold}$ produces a linear decrease of output power, whereas an initial decrease of $T_\mathrm{hot}$ does not affect it.

As a third step, we revisit the prestrain that was already the starting parameter of the load-free analysis. Prestrain is even more decisive for output power, and the trend is sharper (Fig.~\ref{fig:Outputpower}d). A prestrain of at least 2\,\% is required for the TEH to deliver any power, and above $\epsilon_0 \approx 3.2$\,\% the output saturates at $\bar{P} = 190$\,mW. Because both temperatures are kept constant in this series, $\dot{Q}_\mathrm{in}$ is also constant, and the system efficiency relative to Carnot can be read directly off the right-hand axis. The TEH reaches $\eta_\mathrm{sys}/\eta_\mathrm{Carnot} = 3.3 \times 10^{-2}$\,\% as a saturation value rather than a peak, reproduced across the prestrain series and stable from cycle to cycle within each time trace. We return to this efficiency in the discussion.

As an additional step, we examined the influence of fluid flow, completing the set of process parameters available for a TEH. Optimum performance was reached at the maximum water flow our equipment can deliver; the corresponding figure is provided only in the supplementary information, Fig.~S2.

This section delivers both a process-parameter map for maximum output power
and a quantitative picture of how the SMA responds while the device is doing work.
Both follow from the direct measurement of mechanical output,
without using material-property estimates.

\section*{Benchmarking and discussion}

Two figures of merit decide which technology is competitive for low-grade waste-heat recovery: the power per active-material volume, and the power per material cost. We benchmark our TEH against the other ferroic harvesters and against thermoelectric generators (TEG), the most mature technology in this temperature regime. As a first criterion we use the output power scaled to the volume of active material used (Fig.~\ref{fig:Benchmarking}a); an unscaled version is available in supplementary Fig.~S3, but the existing harvesters span a wide range of sizes and an unscaled comparison would not be fair.
Among the thermoelastic harvesters our system reaches the highest power density, 366\,mW/cm³, about 1.7 times that of the next-best TEH demonstrator.\cite{SAKUMA1998,Sato2008,Kumar2019} Next comes the only pyroelectric generator (PEG) realized so far,\cite{Lheritier2022} followed by several thermomagnetic generators (TMG),\cite{Waske2019,Liu2023,Bahl2024} which reach much lower power densities. Across the ferroic-harvester families, only the TEH outperforms TEG, as shown on the right of Fig.~\ref{fig:Benchmarking}a. For this comparison we used only publications in which the output power was directly measured. Many studies report only estimates derived from material properties; while such estimates can identify promising ferroic materials, they cannot be used to compare actual harvester performance. The same restriction substantially reduced the number of TEG demonstrators available for comparison: most TEG papers report only the figure of merit $ZT$, which yields a material efficiency,\cite{Snyder2008} but neglects the junction resistance of the assembled device, which can substantially reduce real system performance. We take 140\,°C as the upper limit of low-grade waste heat in this benchmark, just enough to include one well-characterized TEG\cite{Casi2021} as the strongest competitive reference.
Thermomagnetic microsystems have reached power densities up to 118\,mW/cm³.\cite{Gueltig2017} We exclude these from the comparison since microtechnology is not suited to harvesting the bulk of low-grade waste heat; it addresses a different emerging market, powering the internet of things. Microsystems nevertheless point to the route for raising the output of bulk ferroic harvesters: increasing the cycle frequency.

As a second criterion we use the ratio of output power to material cost, based on current market prices of the raw elements required.\cite{metals} This figure is a first-order proxy for technology cost and often decides whether a technology reaches the market. Given the still low technology-readiness level of all ferroic harvesting demonstrators, a more detailed cost comparison is not yet possible, but materials cost is likely to dominate: all systems require only simple parts for fluid management, and low-grade waste heat itself is supplied nearly for free. The cost ranking in Fig.~\ref{fig:Benchmarking}b broadly reflects the power-density ranking within each technology family, since both scale with the work extracted per cycle from a given active volume; across families, differences in raw-material cost shift the relative positions. Future system optimization should therefore target both axes simultaneously. Our TEH reaches 4.9\,W/€, outperforming every thermoelectric generator benchmarked here.\cite{Casi2021}
For context, photovoltaics, the lowest-cost electricity technology today, delivers about 1\,W/€ at peak,\cite{FraunhoferISE2026}
but this figure includes the full installed system. The 4.9\,W/€ we report for the TEH covers only the SMA material itself; a full-system cost comparison is premature for a lab-scale prototype with custom-built parts and is left to future work. TEH is nevertheless a rapidly emerging technology, as seen in the steep rise of both power density and power-to-price ratio with publication year (Fig.~\ref{fig:Benchmarking}).
The two benchmark numbers therefore reflect two things at once: the system-design choices made in this device, and the material maturity of the SMA at its core. Both can be improved, and the device-material coupling traced through the preceding parameter sweeps shows where the leverage lies.

\begin{figure}[H]
	\centering
	\includegraphics[width=1\linewidth]{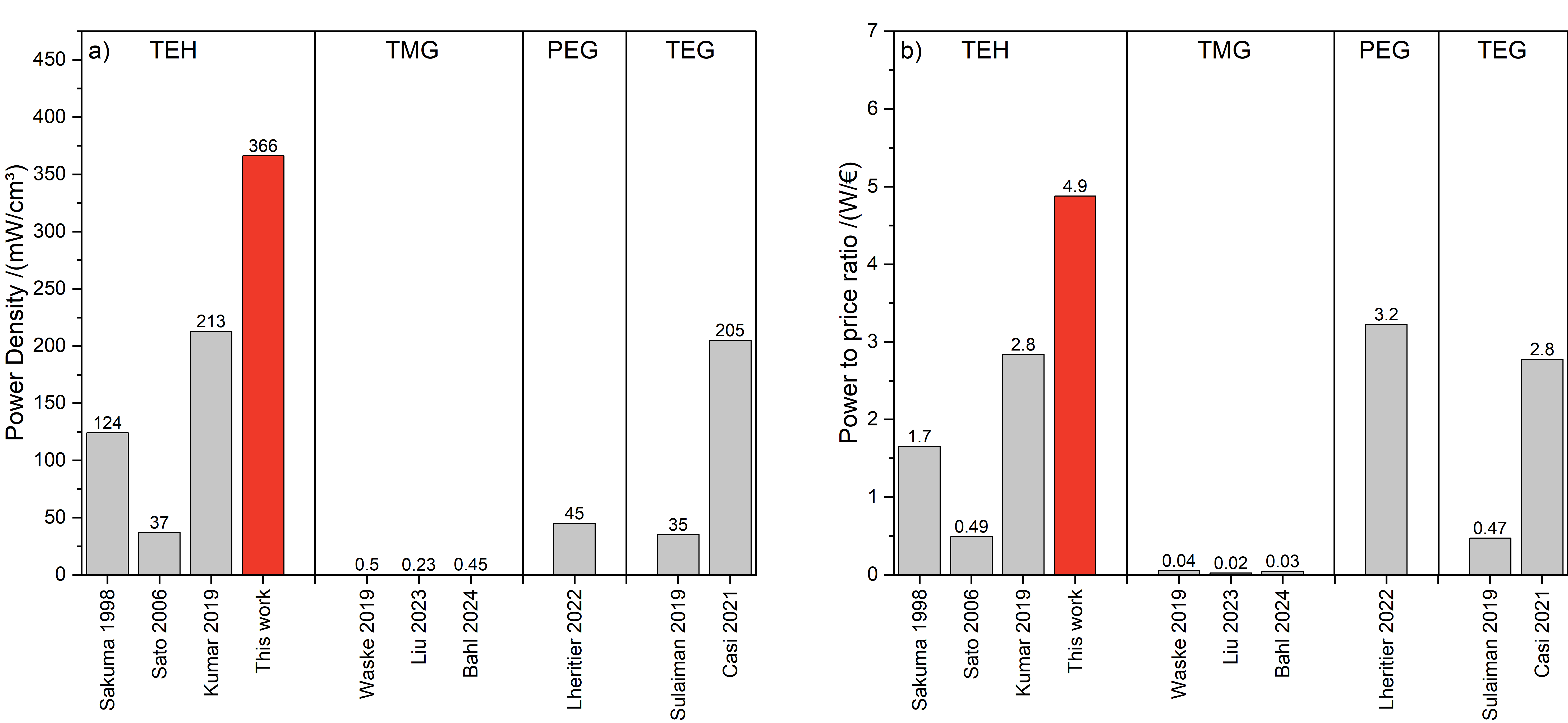}
	\caption[Comparison of power densities and power-to-price ratio of different solid-state low-grade waste-heat harvesting systems]{\textbf{Benchmarking thermoelastic harvester (TEH), pyroelectric generators (PEG), thermomagnetic generators (TMG), and thermoelectric generators (TEG) for harvesting low-grade waste heat.}
       a) Power density relative to the volume of active material required. b) Output power per unit cost of raw functional material.}
	\label{fig:Benchmarking}
\end{figure}

\section*{Conclusion and Outlook}

We have demonstrated thermoelastic harvesting of low-grade waste heat
in a setup that controls every process parameter, including the prestrain,
and measures the output power directly.
On both power density and material-cost-normalized power, our TEH outperforms
every previously reported TEH, thermomagnetic generator and pyroelectric generator,
and outperforms the thermoelectric generators benchmarked in the low-temperature regime.
Beyond these benchmark numbers,
the systematic parameter sweeps deliver a quantitative picture
of how the SMA responds while the device is doing work;
the resulting device--material interaction
will guide the further development of this emerging technology.
The transverse-flow architecture we introduced previously\cite{Neumann2025a}
decouples cycle frequency from wire length,
so output power should rise proportionally with both the number and the length of SMA wires;
the design rules established here transfer directly to such scaled-up systems,
and the next bottleneck lies on the materials side.

To gauge the headroom for improvement,
we compare our measured system efficiency to a material-side simulation of the same SMA at the operating temperatures.\cite{Neumann2025}
The simulation reaches 51.6\,\% of the Carnot bound,
essentially at the Curzon--Ahlborn ceiling for maximum-power operation ($\sim 52$\,\% of Carnot at this $\Delta T$).\cite{Curzon1975}
The material therefore already delivers close to what thermodynamics allows at maximum power;
further gains will come from heat-exchange engineering rather than from the SMA itself,
since at the maximum water-flow optimum most of the supplied hot water exits the chamber warm and unabsorbed.
The material efficiency also bounds the system efficiency,
and across ferroic routes the best thermomagnetic materials reach $\sim 5$\,\%,\cite{Dzekan2021}
consistent with TEH outperforming TMG on power density (Fig.~\ref{fig:Benchmarking}).
TMGs do, however, operate down to temperature spans of 30\,K and below,
whereas our TEH performs best at $\Delta T = 66$\,K.
We attribute the difference to the order of the underlying ferroic phase transitions:
thermoelastic transitions are first-order,
with a hysteresis of several tens of Kelvin that a TEH must overcome,
whereas the best thermomagnetic materials sit at the border between first and second order
and have negligible hysteresis,
which makes TMGs better suited to small temperature spans.

The wires used here are an off-the-shelf NiTi developed for medical applications,
and the system-design rules and benchmarks established in this paper
take this material as a fixed input.
The complementary material-design problem for thermoelastic harvesting
remains largely open:
the harvesting objective is distinct from both the actuator and the elastocaloric cases,
demanding an alloy that maximizes the mechanical work per cycle
while minimizing the latent and sensible heat needed to drive the transformation,
the inverse of the elastocaloric goal of maximizing the heat per stress cycle.
The analogous material-design argument has been laid out for thermomagnetic harvesting,\cite{Dzekan2021}
and design rules for NiTi tailored to elastocaloric cooling
are beginning to emerge.\cite{Frenzel2018,Wieczorek2017}
A dedicated material-design study for thermoelastic harvesting will close the loop,
and the system-design rules and benchmarks established here
provide the figures of merit against which such an alloy should be measured.

\section*{Methods}

\textbf{Wires and material data.} The thermoelastic harvester uses four commercially available Ni$_{56.11}$Ti$_{43.89}$ wires (Fort Wayne Metals) of 1\,mm diameter and 165\,mm active length between the clamps. The wires are used as received, without any thermal or mechanical pre-treatment. Their transformation temperatures are taken from the manufacturer's data sheet: $M_\mathrm{f}=-32\,^\circ$C, $M_\mathrm{s}=-12\,^\circ$C, $A_\mathrm{s}=-8\,^\circ$C and $A_\mathrm{f}=12\,^\circ$C. The density $\rho_\mathrm{NiTi}=6450$\,kg/m³ and the specific heat capacity $c_{p,\mathrm{NiTi}}=340$\,J/(kg\,K) are taken from the supplier data sheet. Two wires are clamped on each side of a symmetric seesaw with stainless-steel holders, each wire fixed by four M3 screws, to form the protagonist--antagonist configuration of Fig.~\ref{fig:Setup}c.

\textbf{Fluid circuit and thermal control.} Each wire bundle is housed in a 3D-printed transversal-flow chamber introduced in \cite{Neumann2025a}. Hot and cold water are supplied by two thermostats (IKA CBC5-Control and RC5-Control), each with an integrated pump, and switched between the two bundles by solenoid valves (Bürkert Type~6628) at the cycle frequency $f$. The lower limit of $T_\mathrm{cold}=7\,^\circ$C is set by the cooling capability of the chiller thermostat. Water volume flows are measured upstream of the chambers by paddle-wheel sensors (Bürkert Type~8031), inlet temperatures by type-K thermocouples (\O\,0.25\,mm), and water and hydraulic pressures by Bürkert Type~8316 transmitters (0--1.5\,bar for water, 0--10\,bar for oil). The thermal input is $\dot{Q}_\mathrm{in}=\dot{V}\,\rho_\mathrm{water}\,c_{p,\mathrm{water}}\,\Delta T$ with $\Delta T=T_\mathrm{hot}-T_\mathrm{cold}$, using literature values of $\rho_\mathrm{water}=998$\,kg/m³ and $c_{p,\mathrm{water}}=4186$\,J/(kg\,K). With $T_\mathrm{hot}=73.7\,^\circ$C and $T_\mathrm{cold}=7.5\,^\circ$C at the point of maximum mechanical output power, and the maximum flow rate of $\dot{V}\approx\SI{0.65}{\litre\per\minute}$, the thermal input power is 2996\,W.

\textbf{Comparing the thermal input with the hydraulic input power through a flowing fluid.} The upper limit of the hydraulic input power can be estimated by taking the maximum input pressure of $p_\mathrm{in} = 0.5$\,bar and assuming that the outflow pressure is zero. With the previously given flow-rate the hydraulic input power equals 1.08\,W. Therefore, the ratio of hydraulic to thermal input is around 0.036\,\%. A schematic of the water-supply system is given in Supplementary Fig.~S4.

\textbf{Mechanical load and prestrain.} Mechanical work is extracted by two oil-filled hydraulic cylinders (Festo DSNU-12-100-P-A, HLP~10 oil with kinematic viscosity 10\,mm²/s), with the load continuously varied by a proportional throttle valve (Bürkert Type~2865). The throttle opening is reported in arbitrary units corresponding to the digital control input to the valve; the absolute oil flow is not calibrated. The prestrain $\epsilon_0$ is set by translating the central bearing with a NEMA-23 stepper motor (47:1 gearbox).

\textbf{Sensors and data acquisition.} Forces on the two sides of the seesaw are measured with Baumer DLM40-IN.502.TP3.C4 sensors, and the seesaw displacement with a Micro-Epsilon optoNCDT~1320 laser triangulation sensor. All signals are acquired at 100\,Hz with two NI~USB-6002 modules under LabVIEW control. The displacement and force signals are smoothed with a Savitzky--Golay filter (polynomial order 3, window length 20 samples) before further processing; raw signals are retained for verification. A complete list of instruments, models, and acquisition settings is given in Supplementary Table~S1.

\textbf{Data analysis.} The first 15\,s of every run are discarded to ensure cyclic steady state. The instantaneous mechanical power on each side is $P_i(t)=F_i(t)\,\dot{x}(t)$, and the output power is the time average of their sum, $\bar{P}=(1/T_\mathrm{m})\int_0^{T_\mathrm{m}}[P_1(t)+P_2(t)]\,dt$, where $T_\mathrm{m}$ is the measurement window. Measured forces and displacements are converted to engineering stress and strain on the SMA using the wire cross-section $A=\pi d^2/4$ with $d=1$\,mm and the active wire length $L=165$\,mm between both clamps: the strain is $\varepsilon=\Delta x/L$ and the stress $\sigma=F/A$, with $\Delta x$ the seesaw displacement at the wire position. The raw system efficiency is $\eta_\mathrm{sys}=\bar{P}/\dot{Q}_\mathrm{in}$, reported throughout the paper as a fraction of the Carnot limit, $\eta_\mathrm{sys}/\eta_\mathrm{Carnot} = \eta_\mathrm{sys}/(1 - T_\mathrm{cold}/T_\mathrm{hot})$. The efficiency is therefore obtained without any assumption on material efficiency or sub-component efficiencies. For the power-density comparison in Fig.~\ref{fig:Benchmarking}a, the volume of ferroic material is the total SMA volume of the four wires, $V_\mathrm{SMA}=4\,(\pi d^2/4)\,L$. For the materials-cost comparison in Fig.~\ref{fig:Benchmarking}b, the cost of each benchmarked device is computed from its reported composition and the reported mass, or, where only the volume is given, from the volume and the material density. The composition is converted to mass fractions, which multiplied by the total mass give the mass of each element. Multiplying each element mass by its raw-material market price\cite{metals} and summing over elements yields the total material cost; the ratio of reported output power to this cost gives the value plotted in Fig.~\ref{fig:Benchmarking}b. Compositions and per-material cost breakdowns are given in Supplementary Table~S3, and the source data and computed values for all benchmarked devices are tabulated in Supplementary Table~S2.

\section*{Acknowledgements}
The authors thank Razie Mohamadi and Ali Izadi for discussions, and
Deutsche Forschungsgemeinschaft (DFG, German Research Foundation) for partly funding this work through grant 549600725.

\section*{Author contributions}
\textbf{B. Neumann}: Conceptualisation (equal); Data curation (lead); Formal analysis (lead); Investigation (lead); Methodology (equal); Software (equal); Validation (equal); Visualisation (lead); Writing -- original draft (lead); Writing -- review \& editing (equal), \textbf{A. Henschke}: Methodology (equal); Resources (equal); Software (lead); \textbf{M. Nikolic}: Data curation (equal); Investigation (equal); Validation (equal), \textbf{S. Fähler}: Conceptualisation (equal); Methodology (equal); Validation (equal); Visualisation (equal); Writing  (lead); Funding (lead).

\section*{Competing interests}
B.N. and S.F. are inventors on German patents DE 10 2023 208 685.9 (transversal fluid management) and DE 10 2020 118 363.1 (prestrain mechanism), both assigned to Helmholtz-Zentrum Dresden--Rossendorf and related to the thermoelastic harvester described in this work. The other authors declare no competing interests.

\section*{Data availability}
The research data of this work is published in the open access data repository RODARE through doi.org/10.14278/rodare.4473.

\clearpage
\setcounter{figure}{0}
\setcounter{table}{0}
\renewcommand{\thefigure}{S\arabic{figure}}
\renewcommand{\thetable}{S\arabic{table}}
\newcounter{suppnote}
\newcommand{\suppnote}[1]{%
  \refstepcounter{suppnote}%
  \vspace{4pt}%
  \subsection*{Supplementary Note \arabic{suppnote}: #1}%
}
\section*{Supplementary Information}

\noindent
This Supplementary Information complements the main manuscript by
providing: (i)~the complete list of sensors, actuators and acquisition
settings (Supplementary~Table~S1);
(ii)~a representative measurement at the optimum operating point, from
which the time-averaged mechanical output power is obtained
(Supplementary~Fig.~S1); (iii)~the influence of the fluid flow rate on
power and efficiency (Supplementary~Fig.~S2); (iv)~an unscaled
(per-system) version of the power benchmark that complements main-text
Fig.~4a (Supplementary~Fig.~S3), together with the source data underlying main-text Fig.~4 (Supplementary~Table~S2);
(v)~a schematic of the hot/cold water-flow circuit
(Supplementary~Fig.~S4); (vi)~the raw-material cost data used for the
cost-normalization comparison (Supplementary~Table~S3); and (vii)~a
video of the operating harvester (Movie~S1, separate file).

\suppnote{Sensors and instrumentation}
\label{sec:instrumentation}

The force with which the seesaw acts on each cylinder is measured by
Baumer DLM40-IN.502.TP3.C4 force sensors mounted below the seesaw and
the cylinder. The seesaw displacement is measured on one side by a
Micro-Epsilon optoNCDT~1320 laser triangulation sensor. The water
temperature is measured at the inlet of each fluid chamber by type-K
thermocouples of \SI{0.25}{mm} diameter. The pressures of the water
flow and the hydraulic oil are measured by B\"urkert 8316 pressure
transmitters (range 0--\SI{1.5}{bar} for water; 0--\SI{10}{bar} for
oil). Fluid flow is measured by B\"urkert Type~8031 low-flow paddle-wheel
sensors, which output a pulse train proportional to the flow rate.
Mechanical work is extracted by two oil-filled hydraulic cylinders (Festo
DSNU-12-100-P-A) acting on a closed hydraulic circuit filled with HLP\,10
oil (kinematic viscosity \SI{10}{\milli\metre\squared\per\second}); the
oil flow through a proportional valve (B\"urkert Type~2865) sets the
mechanical damping continuously and its opening is reported in arbitrary
units (a.u.) corresponding to the digital valve-control value, with
$b=60$\,a.u.\ marking the optimum operating point used in the main text.
All sensor channels are acquired at \SI{100}{\hertz} through two NI
USB-6002 modules under LabVIEW control; cables are shielded and augmented
with small capacitors to suppress noise. All sensors, actuators, and
acquisition settings are summarized in
Supplementary~Table~\ref{tab:teg_sensors}.

\begin{table}[H]
	\centering
	\caption{\textbf{Sensors, actuators, and acquisition settings for
	the thermoelastic harvester measurements.}}
	\label{tab:teg_sensors}
	\setlength{\tabcolsep}{6pt}
	\renewcommand{\arraystretch}{1.2}
	\begin{tabular}{|p{0.18\linewidth}|p{0.26\linewidth}|p{0.20\linewidth}|p{0.30\linewidth}|}
		\hline
		\textbf{Quantity} & \textbf{Device / Model} & \textbf{Output \& range} & \textbf{Note} \\
		\hline
		Flow (hot/cold) &
		B\"urkert Type 8031 (paddle wheel) &
		Frequency / pulses, 0/\SI{10}{\volt} &
		Inline, upstream of chambers; used for $\dot{V}$ and heat-input
		calculation \\
		\hline
		Force ($\times 2$) &
		Baumer DLM40-IN.502.TP3.C4 &
		\SIrange{0}{10}{\volt} &
		Mounted below seesaw and hydraulic cylinder \\
		\hline
		Displacement &
		Micro-Epsilon optoNCDT~1320 (laser triangulation) &
		\SIrange{4}{20}{\milli\ampere} &
		Measurement of the seesaw motion on one side \\
		\hline
		Temperature (inlet) &
		Type-K thermocouple, $\varnothing\,$\SI{0.25}{mm} &
		\SIrange{0}{5}{\volt} $\rightarrow$ NI-6002 AI &
		At chamber inlets; used to compute
		$\Delta T = T_\mathrm{hot}-T_\mathrm{cold}$ \\
		\hline
		Pressure &
		B\"urkert 8316 (pressure transmitter) &
		\SIrange{4}{20}{\milli\ampere} &
		Pressure of fluid flow and hydraulic oil (water: 0--\SI{1.5}{bar};
		oil: 0--\SI{10}{bar}) \\
		\hline
		Switch valve &
		B\"urkert 6628 (solenoid switch valve) &
		\SI{24}{\volt} digital &
		Switches the water flow between bundles at the cycle frequency
		$f$ \\
		\hline
		Variable damping &
		B\"urkert Type 2865 (proportional valve controller) &
		\SIrange{0}{10}{\volt} control &
		Sets the hydraulic damping (reported in a.u.\ corresponding to
		the digital control input) \\
		\hline
		Mechanical load &
		Festo DSNU-12-100-P-A (cylinders) &
		--- &
		Linear cylinders to convert mechanical to hydraulic work \\
		\hline
		Working fluid (oil) &
		Hydraulic oil HLP\,10 &
		Viscosity \SI{10}{\milli\metre\squared\per\second} &
		Used in the hydraulic damping circuit \\
		\hline
		Thermostat (hot) &
		IKA CBC5-Control&
		Set point $\leq\SI{85}{\degreeCelsius}$ &
		Supplies hot fluid; pump in built-in circuit \\
		\hline
		Thermostat (cold) &
		IKA RC5-Control&
		Set point $\geq\SI{5}{\degreeCelsius}$ &
		Supplies cold fluid \\
		\hline
		Prestrain drive &
		NEMA~23 stepper motor with 47:1 gearbox &
		\SI{24}{\volt}, \SI{2.8}{\ampere} &
		Translates the central bearing to set $\epsilon_0$ \\
		\hline
		DAQ \& control &
		NI USB-6002 ($\times 2$) &
		16-bit, \SI{100}{\hertz} &
		LabVIEW logging to CSV \\
		\hline
	\end{tabular}
\end{table}

\suppnote{Power output at the optimum operating point}
\label{sec:load}

With the hydraulic cylinders attached, the harvester performs
mechanical work against a tunable damper. The damping is reported in
arbitrary units (a.u.), corresponding to the digital control input to
the proportional valve. Each working point in main-text Fig.~3 was
identified by jointly sweeping switching frequency~$f$ and damping~$b$.

\begin{figure}[H]
	\centering
	\includegraphics[width=1\linewidth]{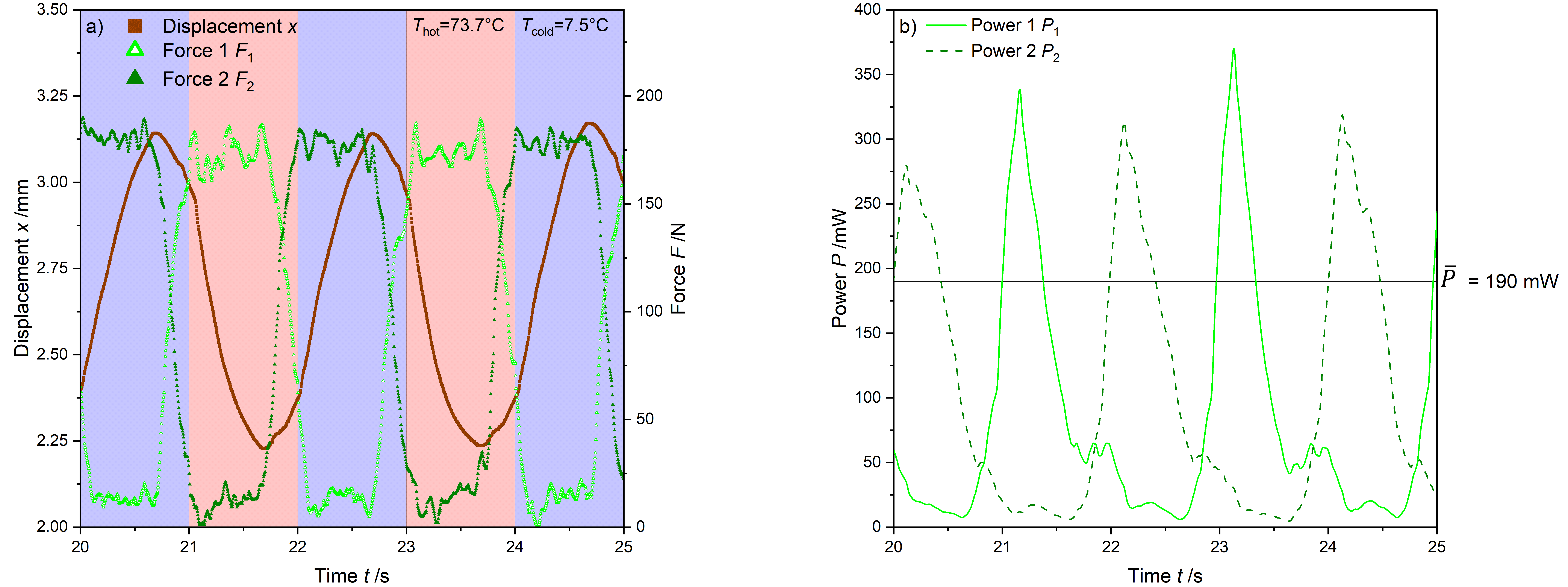}
	\caption{\textbf{Determining mechanical output power under load
	from displacement and force measurements.}
	\textbf{a)}~When the fluid flow alternates between hot and cold,
	both sets of wires generate alternating tensile forces $F_1$ and
	$F_2$ that accelerate the seesaw back and forth. Because work is
	extracted on both half-cycles in the protagonist-antagonist
	configuration, a force peak is measured in each movement direction.
	\textbf{b)}~Instantaneous power $P_i=F_i\dot{x}$ on the two sides
	together with the time-averaged total power $\bar{P}$. The peak
	mechanical power is \SI{370}{\milli\watt} and the average is
	\SI{190}{\milli\watt}. Conditions: $f=\SI{1}{\hertz}$,
	$\epsilon_0=\SI{3.6}{\percent}$,
	$T_\mathrm{cold}=\SI{7.5}{\degreeCelsius}$,
	$T_\mathrm{hot}=\SI{73.7}{\degreeCelsius}$, $b=60$~a.u.}
	\label{fig:figureforoptimalcondi}
\end{figure}

\paragraph{Computing the average power from force and displacement.}
Supplementary~Fig.~\ref{fig:figureforoptimalcondi} shows a
representative measurement at the conditions that maximize the output
power, and illustrates how the time-averaged mechanical power is
obtained from the measured signals. The forces on both sides have a
near-rectangular shape in each half-cycle, indicating a near-constant force during the entire heating step. The displacement follows
a quasi-sinusoidal trajectory and is phase-shifted by about $90^\circ$
relative to the force, as expected at the optimum mechanical load. The
instantaneous power on each side is $P_i = F_i \dot{x}$, and the
reported output power is the time average of their sum,
$\bar{P} = \tfrac{1}{T_\mathrm{m}}\int_0^{T_\mathrm{m}}
[P_1(t)+P_2(t)]\,\mathrm{d}t$. The peak instantaneous power is
\SI{370}{\milli\watt}; the time-averaged mechanical power is
$\bar{P}=\SI{190}{\milli\watt}$.

\suppnote{Influence of the fluid flow rate}
\label{sec:flowsweep}

Because lowering the hot-side temperature substantially improved
efficiency, the fluid flow rate, which together with $\Delta T$
sets the thermal input power,
\begin{equation}
\dot{Q}_\mathrm{in} = \dot{V}\,\rho\,c_p\,\Delta T,
\qquad \Delta T = T_\mathrm{hot}-T_\mathrm{cold},
\label{eq:Qin}
\end{equation}
was also varied. The flow rate was changed via the rotational speed of
the thermostat pumps. As shown in
Supplementary~Fig.~\ref{fig:effvsflow}, both average power and system
efficiency decrease with decreasing flow rate. The flow
rate was therefore left at the maximum value the thermostats could
deliver, $\dot{V}\approx\SI{0.65}{\litre\per\minute}$, for all other
experiments reported in this work.

\begin{figure}[H]
	\centering

	\includegraphics[width=0.7\linewidth]{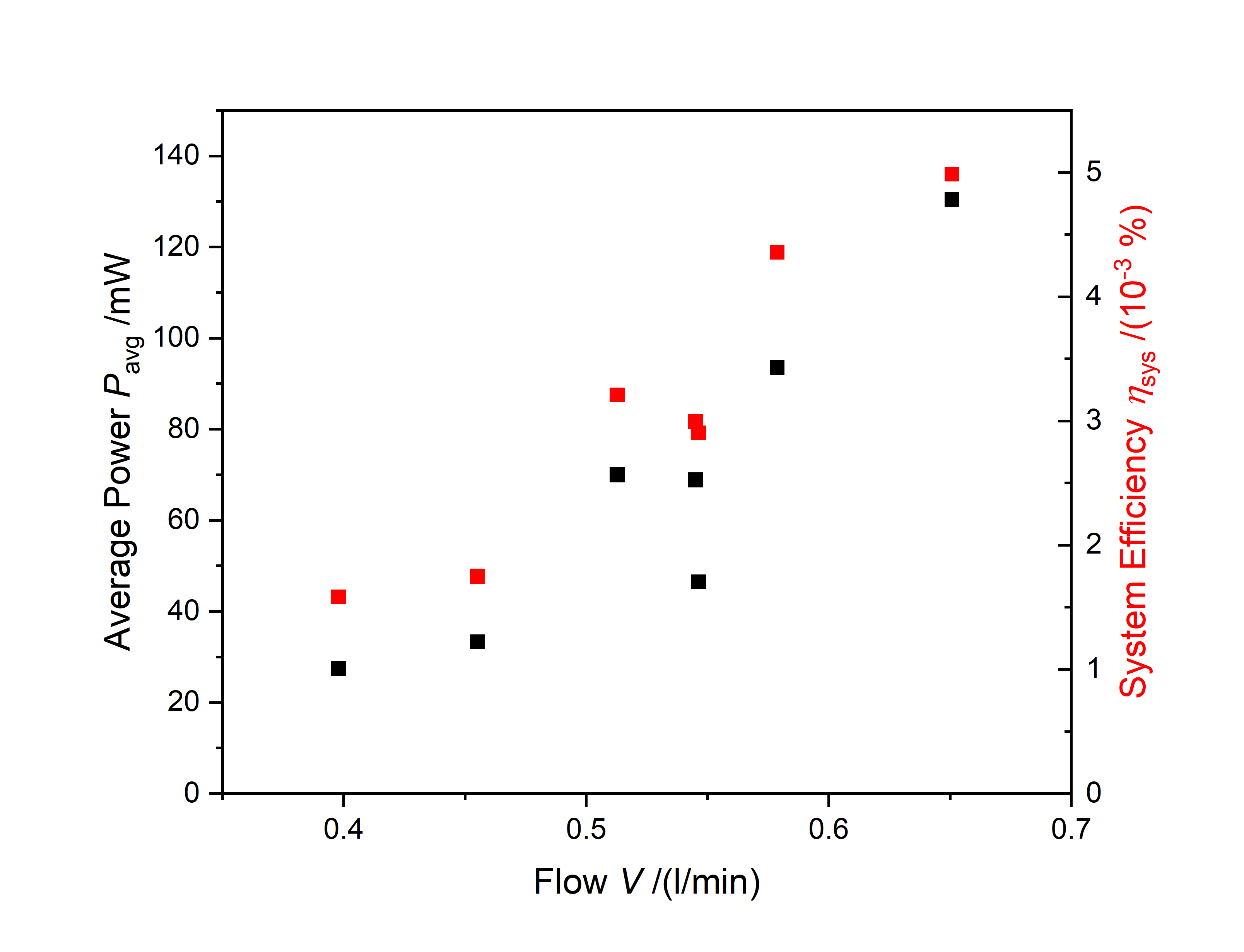}
	\caption{\textbf{Influence of the water flow rate on the average
	power and system efficiency.} Both quantities decrease
	monotonically with decreasing flow rate, so the maximum flow rate
	the thermostats could deliver
	($\dot{V}\approx\SI{0.65}{\litre\per\minute}$) was used in all
	other experiments.}
	\label{fig:effvsflow}
\end{figure}

\suppnote{Benchmarking: unscaled total output power}
\label{sec:totalpower}

\begin{figure}[H]
	\centering
	\includegraphics[width=0.85\linewidth]{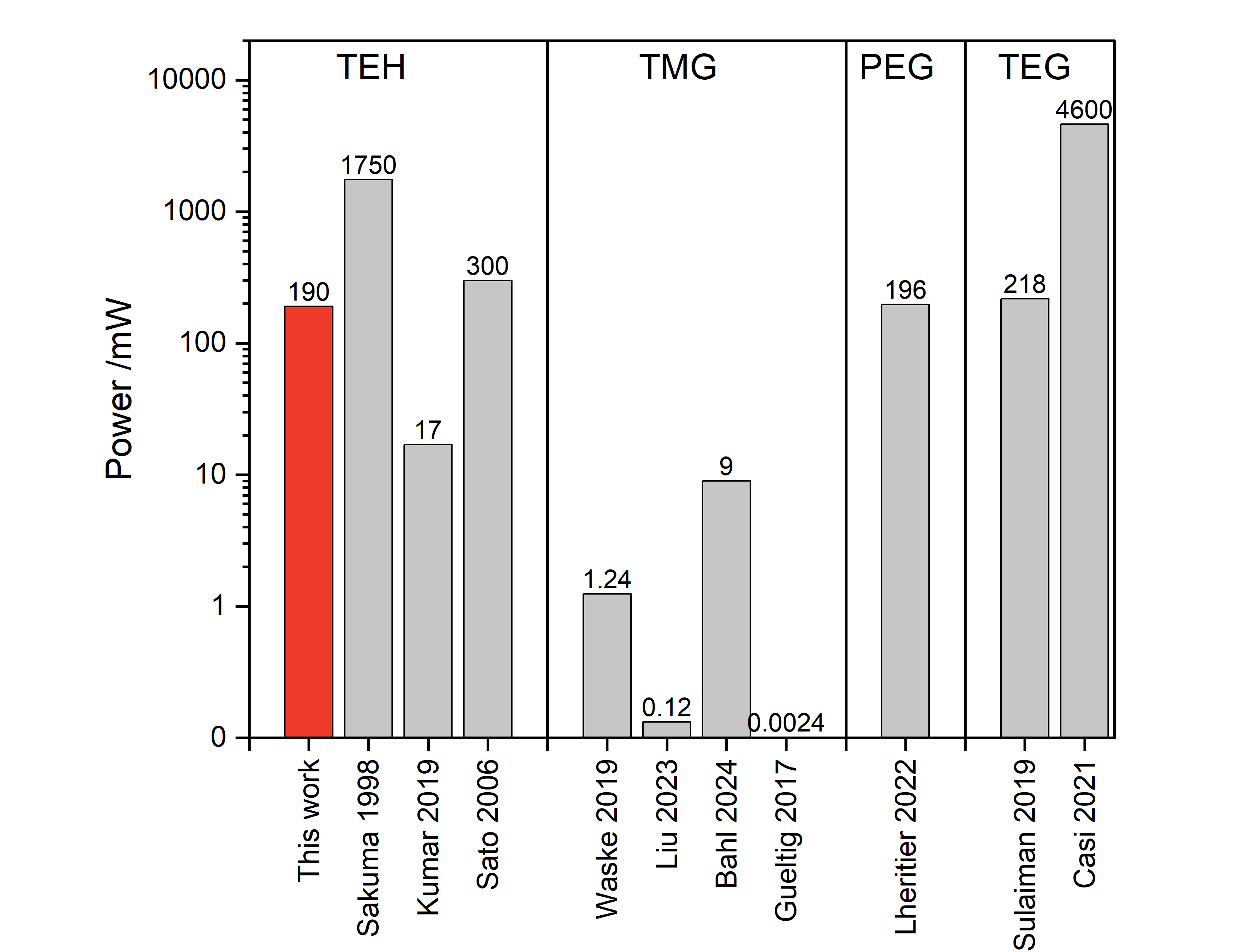}
	\caption{\textbf{Total output power of solid-state low-grade
	waste-heat harvesting systems, unscaled by active-material volume.}
	Bars give the measured electrical (TMG, PEG, TEG) or mechanical
	(TEH) output power. All entries refer to systems operating below
	\SI{140}{\degreeCelsius} with directly measured (not estimated)
	power values. References for each entry are those used in
	main-text Fig.~4.}
	\label{fig:totalpower}
\end{figure}

The main-text Fig.~4a benchmarks the harvesters by power density
(per active-material volume) and Fig.~4b by power per material cost.
For completeness, the same systems are compared here on the basis of
\emph{total} output power, without scaling to active-material volume
(Supplementary~Fig.~\ref{fig:totalpower}). As pointed out in the main
text, this comparison is less informative because the different
demonstrators contain very different amounts of active material. The
systems of Sato and Sakuma\cite{Sato2008,SAKUMA1998} use bulky wires
and therefore reach higher total power than the protagonist-antagonist device reported here, while their power
densities are substantially lower (main-text Fig.~4a). The same caveat
applies to the largest thermoelectric demonstrators on the right of
the figure. The unscaled comparison shown here is therefore best read
as a complement to the volume- and cost-normalized benchmarks of the
main text, not as a substitute. All entries refer to systems
operating with waste-heat temperatures $\le\SI{140}{\degreeCelsius}$
and report directly measured output power (not estimates from
material properties). The source data underlying main-text Fig.~4 are
listed in Supplementary~Table~S2.

\begin{table}[H]
	\centering
	\caption{\textbf{Source data for the benchmarking comparison in
	main-text Fig.~4.} For every benchmarked TEH, PEG, TMG and TEG
	device: the reported output power, the active-material volume (or
	mass), the resulting power density, the material cost computed from
	composition and raw-material prices, and the cost-normalized power.}
	\label{tab:benchmark_data}
	\resizebox{\textwidth}{!}{%
	\begin{tabular}{l l l
			S[table-format=4.1]
			S[table-format=2.3]
			S[table-format=3.0]
			S[table-format=2.2]
			S[table-format=1.2]}
		\toprule
		Class & Device (year) & Ref. &
		{$P$ (mW)} & {$V$ (cm$^3$)} &
		{Power dens.\ (mW/cm$^3$)} & {Cost (\,€)} & {$P$/cost (W/\,€)} \\
		\midrule
		TEH & Sakuma 1998 & \cite{SAKUMA1998} & {1750} & {14.11} & 124  & {1.06} & 1.65 \\
		TEH & Sato 2008   & \cite{Sato2008}   & {300}  & {8.1}   &  37  & {0.61} & 0.49 \\
		TEH & Kumar 2019  & \cite{Kumar2019}  & {17}   & {0.08}  & 213  & {0.006} & 2.84 \\
		TEH & This work   & ---               & {190}  & {0.518} & 366  & {0.039} & 4.88 \\
		\midrule
		TMG & Waske 2019  & \cite{Waske2019}  & {1.24} & {2.48} & 0.5  & {0.023} & 0.05 \\
		TMG & Liu 2023    & \cite{Liu2023}    & {0.12} & {0.52} & 0.23 & {0.005} & 0.02 \\
		TMG & Bahl 2024   & \cite{Bahl2024}   & {9}    & {20}   & 0.45 & {0.188} & 0.05 \\
		\midrule
		PEG & Lheritier 2022 & \cite{Lheritier2022} & {196} & {4.4} & 45  & {0.061} & 3.23 \\
		\midrule
		TEG & Sulaiman 2019 & \cite{SaufiSulaiman2019} & {218}  & {6.22}  &  35 & {0.46} & 0.47 \\
		TEG & Casi 2021    & \cite{Casi2021}           & {4600} & {22.44} & 205 & {1.66} & 2.78 \\
		\bottomrule
	\end{tabular}}
\end{table}

\suppnote{Fluid circuit}
\label{sec:fluidcircuit}

The water supply for the harvester experiments is the same as introduced
by Neumann et al.\cite{Neumann2025a} for the chamber characterization, except that the fluid-flow
sensors were exchanged (see Supplementary~Note~\ref{sec:instrumentation}).
Hot and cold water are circulated by two thermostats (IKA RC5-Control for
the hot side and CBC5-Control for the cold side), and the hot/cold flows
are switched between the two wire bundles by solenoid valves
(B\"urkert~Type~6628) at the cycle frequency~$f$. The full hydraulic
layout is sketched in Supplementary~Fig.~\ref{fig:fluidplan}.

\begin{figure}[H]
	\centering
	\includegraphics[width=0.75\linewidth]{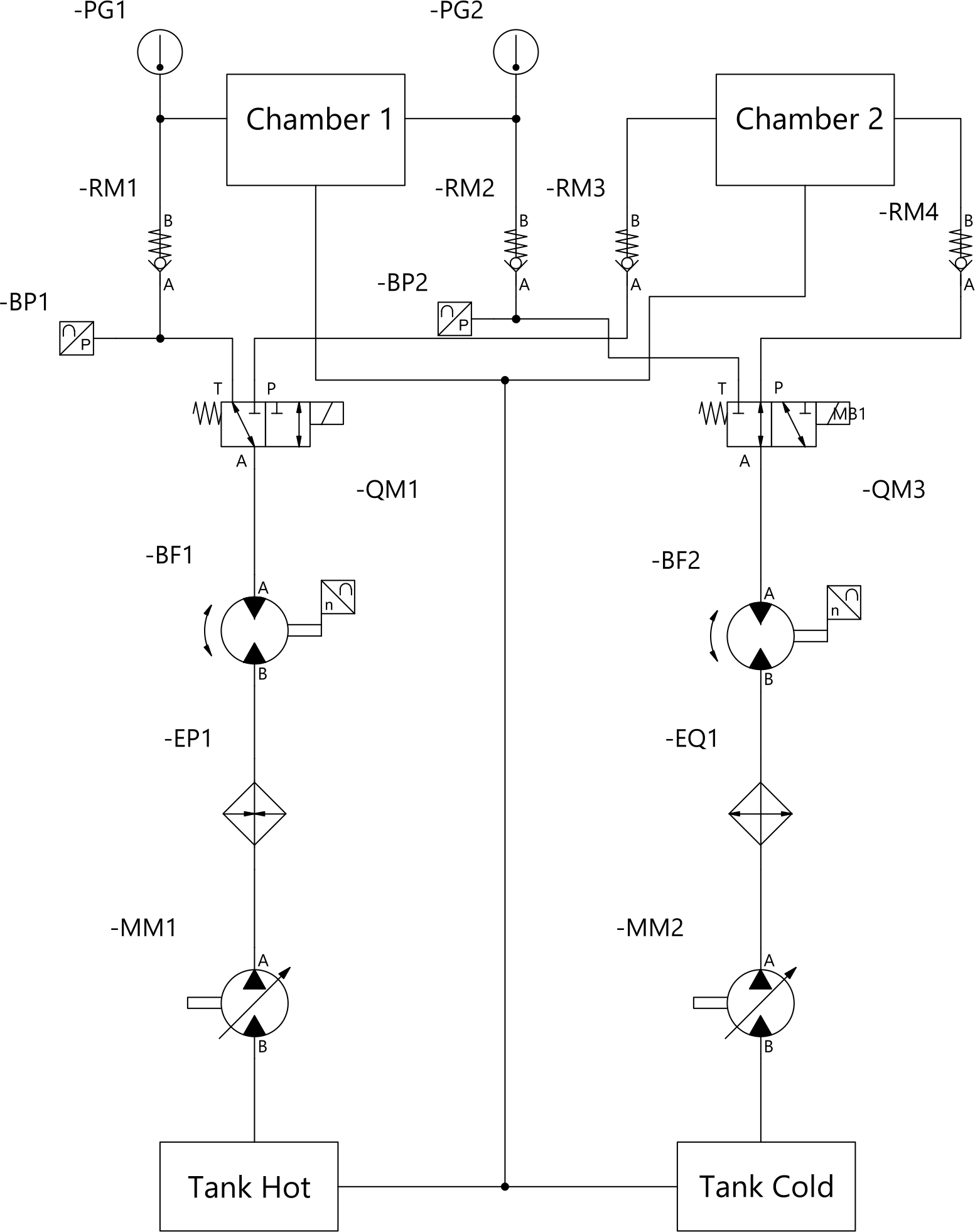}
	\caption{\textbf{Schematic of the hydraulic circuit for hot and cold
	water flow.} Both hot and cold fluid circuits have their own
	reservoir from which pumps (\textbf{MM1, MM2}) generate the required
	pressure and flow. Heater (\textbf{EP1}) and cooler (\textbf{EQ1})
	bring the water to temperature. Downstream, the fluid flow is
	measured (\textbf{BF1, BF2}) and distributed to the chambers by
	magnetic switching valves (\textbf{QM1, QM3}). Before entering the
	chambers, pressure (\textbf{BP1, BP2}) and temperature
	(\textbf{PG1, PG2}) are measured. One-way valves (\textbf{RM1--4})
	prevent back-flow. After heat exchange with the wires, the water
	returns to the corresponding tank.}
	\label{fig:fluidplan}
\end{figure}

\suppnote{Raw-material cost of the active materials}

\begin{table}[H]
	\centering
	\caption[Raw-material cost of the active materials]{\textbf{Raw-material cost of the active materials of different low-grade waste-heat harvesting technologies.} For each active material the constituent base materials are listed with their atomic composition, the raw-material price taken from \texttt{metal.com} (retrieved 6 July 2026), the atomic mass $M$ and the resulting mass-weighted fraction $w_i$. The price per kilogram of active material follows from $\langle p\rangle=\sum_i w_i\,p_i$ and the price per unit volume from $\langle p\rangle$ and the mass density $\rho$. Compositions of LaFeCoSi \cite{Lollobrigida2014} and PST \cite{Lim2002} are taken from the literature. At the time of acquisition of this data, one US-dollar was worth 0.88 Euros.}
	\label{tab:material_cost}
	\resizebox{\textwidth}{!}{%
	\begin{tabular}{l l
			S[table-format=2.2]
			S[table-format=5.0]
			S[table-format=3.2]
			S[table-format=1.3]
			S[table-format=2.2]
			S[table-format=1.4]
			S[table-format=1.2]}
		\toprule
		Active material & Base material & {Comp.\ (at.\%)} & {Price (USD/t)} & {$M$ (u)} & {$w_i$}
		& {$\langle p\rangle$ (USD/kg)} & {$\langle p\rangle$ (USD/cm\textsuperscript{3})} & {$\rho$ (g/cm\textsuperscript{3})} \\
		\midrule
		Nickel--Titanium (NiTi) & Nickel   & 56.11 & 16300 & 58.70 & 0.611 & 13.22 & 0.0853 & 6.45 \\
		                        & Titanium & 43.89 &  8383 & 47.90 & 0.389 &       &        &      \\
		\midrule
		LaFeCoSi & Lanthanum & 7.80 &   809 & 138.91 & 0.181 & 1.35 & 0.0107 & 7.90 \\
		         & Iron      & 81.40 &  639 &  55.80 & 0.760 &      &        &      \\
		         & Cobalt    &  1.50 & 44787 &  58.90 & 0.015 &      &        &      \\
		         & Silicon   &  9.30 &  1272 &  28.10 & 0.044 &      &        &      \\
		\midrule
		PST (lead--scandium--tantalate) & Lead     & 50.00 & 2331 & 207.20 & 0.647 & 1.76 & 0.0159 & 9.00 \\
		                                & Scandium & 25.00 &  845 &  45.00 & 0.070 &      &        &      \\
		                                & Tantalum & 25.00 &  685 & 181.00 & 0.283 &      &        &      \\
		\midrule
		Bismuth--telluride (Bi$_2$Te$_3$) & Bismuth   & 40.00 & 20045 & 209.00 & 0.522 & 10.68 & 0.0839 & 7.85 \\
		                                  & Tellurium & 60.00 &   463 & 127.60 & 0.478 &       &        &      \\
		\bottomrule
	\end{tabular}}
\end{table}

\suppnote{Supplementary Movie}

\noindent
\textbf{Movie~S1.} Video of the operating TEH, showing the
reciprocating motion of the seesaw driven by the alternating
martensite-to-austenite transformation of the two opposing sets of NiTi
wires. The movie is provided as a separate file
(\texttt{Movie\_S1.mov}).

\bibliographystyle{unsrtnat}
\bibliography{literature_library}

\begin{thebibliography}{51}
\providecommand{\natexlab}[1]{#1}
\providecommand{\url}[1]{\texttt{#1}}
\expandafter\ifx\csname urlstyle\endcsname\relax
  \providecommand{\doi}[1]{doi: #1}\else
  \providecommand{\doi}{doi: \begingroup \urlstyle{rm}\Url}\fi

\bibitem[Zhang et~al.(2026)Zhang, Shan, Li, Xue, Ma, Kikstra, Shi, Wang, Zhang,
  Wang, Fang, Yang, and Hubacek]{Zhang2026}
Hongzhi Zhang, Yuli Shan, Ruoqi Li, Rui Xue, Junhua Ma, Jarmo Kikstra, Zongbo
  Shi, Zhaohua Wang, Bin Zhang, Bo~Wang, Shuai Fang, Fan Yang, and Klaus
  Hubacek.
\newblock Rising air-conditioning use intensifies global warming.
\newblock \emph{Nature Communications}, 17\penalty0 (1):\penalty0 1961, 2026.
\newblock \doi{10.1038/s41467-026-69393-1}.

\bibitem[Forman et~al.(2016)Forman, Muritala, Pardemann, and Meyer]{Forman2016}
Clemens Forman, Ibrahim~Kolawole Muritala, Robert Pardemann, and Bernd Meyer.
\newblock Estimating the global waste heat potential.
\newblock \emph{Renewable and Sustainable Energy Reviews}, 57:\penalty0
  1568--1579, 2016.
\newblock \doi{10.1016/j.rser.2015.12.192}.

\bibitem[Liu et~al.(2012)Liu, Gottschall, Skokov, Moore, and
  Gutfleisch]{Liu2012}
Jian Liu, Tino Gottschall, Konstantin~P. Skokov, James~D. Moore, and Oliver
  Gutfleisch.
\newblock Giant magnetocaloric effect driven by structural transitions.
\newblock \emph{Nature Materials}, 11\penalty0 (7):\penalty0 620--626, 2012.
\newblock \doi{10.1038/nmat3334}.

\bibitem[Li et~al.(2023)Li, Torell{\'o}, Kovacova, Prah, Aravindhan, Granzow,
  Usui, Hirose, and Defay]{Li2023}
Junning Li, Alvar Torell{\'o}, Veronika Kovacova, Uros Prah, Ashwath
  Aravindhan, Torsten Granzow, Tomoyasu Usui, Sakyo Hirose, and Emmanuel Defay.
\newblock High cooling performance in a double-loop electrocaloric heat pump.
\newblock \emph{Science}, 382\penalty0 (6673):\penalty0 801--805, 2023.
\newblock \doi{10.1126/science.adi5477}.

\bibitem[Tu{\v{s}}ek et~al.(2016)Tu{\v{s}}ek, Engelbrecht, Eriksen, Dall'Olio,
  Tu{\v{s}}ek, and Pryds]{Tusek2016}
Jaka Tu{\v{s}}ek, Kurt Engelbrecht, Dan Eriksen, Stefano Dall'Olio, Janez
  Tu{\v{s}}ek, and Nini Pryds.
\newblock A regenerative elastocaloric heat pump.
\newblock \emph{Nature Energy}, 1\penalty0 (10):\penalty0 16134, 2016.
\newblock \doi{10.1038/nenergy.2016.134}.

\bibitem[F{\"a}hler et~al.(2012)F{\"a}hler, R{\"o}{\ss}ler, Kastner, Eckert,
  Eggeler, Emmerich, Entel, M{\"u}ller, Quandt, and Albe]{faehler2012}
Sebastian F{\"a}hler, Ulrich~K. R{\"o}{\ss}ler, Oliver Kastner, J{\"u}rgen
  Eckert, Gunther Eggeler, Heike Emmerich, Peter Entel, Stefan M{\"u}ller,
  Eckhard Quandt, and Karsten Albe.
\newblock Caloric effects in ferroic materials: {New} concepts for cooling.
\newblock \emph{Advanced Engineering Materials}, 14\penalty0 (1--2):\penalty0
  10--19, 2012.
\newblock \doi{10.1002/adem.201100178}.

\bibitem[Bahl et~al.(2024)Bahl, Engelbrecht, Gideon, Levy, Marcussen,
  Imbaquingo, and Bj{\o}rk]{Bahl2024}
Christian~R.H. Bahl, Kurt Engelbrecht, Arendse Gideon, Mikael
  Alexander~Vinogradov Levy, Jacob~Birkj{\ae}r Marcussen, Carlos Imbaquingo,
  and Rasmus Bj{\o}rk.
\newblock Design, optimization and operation of a high power thermomagnetic
  harvester.
\newblock \emph{Applied Energy}, 376:\penalty0 124304, 2024.
\newblock \doi{10.1016/j.apenergy.2024.124304}.

\bibitem[Waske et~al.(2019)Waske, Dzekan, Sellschopp, Berger, Stork, Nielsch,
  and F{\"a}hler]{Waske2019}
Anja Waske, Daniel Dzekan, Kai Sellschopp, Dietmar Berger, Alexander Stork,
  Kornelius Nielsch, and Sebastian F{\"a}hler.
\newblock Energy harvesting near room temperature using a thermomagnetic
  generator with a pretzel-like magnetic flux topology.
\newblock \emph{Nature Energy}, 4:\penalty0 68--74, 2019.
\newblock \doi{10.1038/s41560-018-0306-x}.

\bibitem[Gueltig et~al.(2017)Gueltig, Wendler, Ossmer, Ohtsuka, Miki, Takagi,
  and Kohl]{Gueltig2017}
Marcel Gueltig, Frank Wendler, Hinnerk Ossmer, Makoto Ohtsuka, Hiroyuki Miki,
  Toshiyuki Takagi, and Manfred Kohl.
\newblock High-performance thermomagnetic generators based on heusler alloy
  films.
\newblock \emph{Advanced Energy Materials}, 7\penalty0 (5):\penalty0 1601879,
  2017.
\newblock \doi{https://doi.org/10.1002/aenm.201601879}.

\bibitem[Lheritier et~al.(2022)Lheritier, Torell{\'o}, Usui, Nouchokgwe,
  Aravindhan, Li, Prah, Kovacova, Bouton, Hirose, and Defay]{Lheritier2022}
Pierre Lheritier, Alvar Torell{\'o}, Tomoyasu Usui, Youri Nouchokgwe, Ashwath
  Aravindhan, Junning Li, Uros Prah, Veronika Kovacova, Olivier Bouton, Sakyo
  Hirose, and Emmanuel Defay.
\newblock Large harvested energy with non-linear pyroelectric modules.
\newblock \emph{Nature}, 609\penalty0 (7928):\penalty0 718--721, 2022.
\newblock \doi{10.1038/s41586-022-05069-2}.

\bibitem[Goldstein and McNamara(1978)]{LE1978}
David~M. Goldstein and Leo~J. McNamara, editors.
\newblock \emph{Proceedings of the NITINOL Heat Engine Conference}, Silver
  Spring, Maryland, 1978.

\bibitem[Sato et~al.(2008)Sato, Yoshida, Tanabe, Fujita, and Ooiwa]{Sato2008}
Yoshihisa Sato, Naotsugu Yoshida, Yukinori Tanabe, Hideki Fujita, and Norio
  Ooiwa.
\newblock Characteristics of a new power generation system with application of
  a shape memory alloy engine.
\newblock \emph{Electrical Engineering in Japan}, 165\penalty0 (3):\penalty0
  8--15, 2008.
\newblock \doi{https://doi.org/10.1002/eej.20620}.

\bibitem[Kumar et~al.(2019)Kumar, Kishore, Maurya, Stewart, Mirzaeifar, Quandt,
  and Priya]{Kumar2019}
Prashant Kumar, Ravi~Anant Kishore, Deepam Maurya, Colin~J. Stewart, Reza
  Mirzaeifar, Eckhard Quandt, and Shashank Priya.
\newblock Shape memory alloy engine for high efficiency low-temperature
  gradient thermal to electrical conversion.
\newblock \emph{Applied Energy}, 251:\penalty0 113277, 2019.
\newblock \doi{10.1016/j.apenergy.2019.05.080}.

\bibitem[Liang et~al.(2026)Liang, Pickett, Hermann, Sittig, Reichert, Lehmann,
  St{\"o}tzer, Zwick, Greifenstein, Strauch, Skokov, Gutfleisch, Gottschall,
  Fries, and Benke]{Liang2026}
Jierong Liang, Jeffrey Pickett, Sarah Hermann, Tim Sittig, Thomas Reichert,
  Martin Lehmann, Jonas St{\"o}tzer, Jens-Peter Zwick, Max Greifenstein, Milan
  Strauch, Konstantin Skokov, Oliver Gutfleisch, Tino Gottschall, Maximilian
  Fries, and Dimitri Benke.
\newblock Polaris: from laboratory prototypes to market-ready sustainable
  magnetic beverage coolers.
\newblock \emph{Applied Thermal Engineering}, 284:\penalty0 129144, 2026.
\newblock \doi{10.1016/j.applthermaleng.2025.129144}.

\bibitem[{International Energy Agency}(2018)]{IEA2018}
{International Energy Agency}.
\newblock The future of cooling: Opportunities for energy-efficient air
  conditioning.
\newblock Technical report, OECD/IEA, Paris, 2018.
\newblock URL \url{https://www.iea.org/reports/the-future-of-cooling}.

\bibitem[Firth et~al.(2019)Firth, Zhang, and Yang]{Firth2019}
Anton Firth, Bo~Zhang, and Aidong Yang.
\newblock Quantification of global waste heat and its environmental effects.
\newblock \emph{Applied Energy}, 235:\penalty0 1314--1334, 2019.
\newblock \doi{10.1016/j.apenergy.2018.10.102}.

\bibitem[Papapetrou et~al.(2018)Papapetrou, Kosmadakis, Cipollina, {La
  Commare}, and Micale]{Papapetrou2018}
Michael Papapetrou, George Kosmadakis, Andrea Cipollina, Umberto {La Commare},
  and Giorgio Micale.
\newblock Industrial waste heat: Estimation of the technically available
  resource in the eu per industrial sector, temperature level and country.
\newblock \emph{Applied Thermal Engineering}, 138:\penalty0 207--216, 2018.
\newblock \doi{https://doi.org/10.1016/j.applthermaleng.2018.04.043}.

\bibitem[Hao et~al.(2025)Hao, Zhou, Tian, Zhang, Zhou, Shen, Wu, and
  Li]{Hao2025}
Yueting Hao, Haojie Zhou, Tong Tian, Wei Zhang, Xin Zhou, Qingfei Shen, Tong
  Wu, and Ji~Li.
\newblock Data centers waste heat recovery technologies: Review and evaluation.
\newblock \emph{Applied Energy}, 384:\penalty0 125489, 2025.
\newblock \doi{https://doi.org/10.1016/j.apenergy.2025.125489}.

\bibitem[Yuan et~al.(2025)Yuan, Liu, Sun, Lin, Fan, Zhao, and
  Kosonen]{Yuan2025}
Xiaolei Yuan, Jiayi Liu, Sijia Sun, Xiaojie Lin, Xiaojun Fan, Weixin Zhao, and
  Risto Kosonen.
\newblock Data center waste heat for district heating networks: A review.
\newblock \emph{Renewable and Sustainable Energy Reviews}, 219:\penalty0
  115863, 2025.
\newblock \doi{https://doi.org/10.1016/j.rser.2025.115863}.

\bibitem[Jaziri et~al.(2020)Jaziri, Boughamoura, M{\"u}ller, Mezghani, Tounsi,
  and Ismail]{Jaziri2020}
Nesrine Jaziri, Ayda Boughamoura, Jens M{\"u}ller, Brahim Mezghani, Fares
  Tounsi, and Mohammed Ismail.
\newblock A comprehensive review of thermoelectric generators: Technologies and
  common applications.
\newblock \emph{Energy Reports}, 6:\penalty0 264--287, 2020.
\newblock \doi{10.1016/j.egyr.2019.12.011}.

\bibitem[Luo et~al.(2014)Luo, Li, Cai, Zhou, Tang, Zhai, and Zhang]{Luo2014}
Qi~Luo, Peng Li, Lanlan Cai, Pingwang Zhou, Di~Tang, Pengcheng Zhai, and
  Qingjie Zhang.
\newblock A thermoelectric waste-heat-recovery system for portland cement
  rotary kilns.
\newblock \emph{Journal of Electronic Materials}, 44\penalty0 (6):\penalty0
  1750--1762, December 2014.
\newblock \doi{10.1007/s11664-014-3543-1}.

\bibitem[Champier(2017)]{Champier2017}
Daniel Champier.
\newblock Thermoelectric generators: A review of applications.
\newblock \emph{Energy Conversion and Management}, 140:\penalty0 167--181,
  2017.
\newblock \doi{10.1016/j.enconman.2017.02.070}.

\bibitem[Kishore and Priya(2018)]{Kishore2018}
Ravi~Anant Kishore and Shashank Priya.
\newblock A review on low-grade thermal energy harvesting: Materials, methods
  and devices.
\newblock \emph{Materials}, 11\penalty0 (8):\penalty0 1433, 2018.
\newblock \doi{10.3390/ma11081433}.

\bibitem[{Saufi Sulaiman} et~al.(2019){Saufi Sulaiman}, Singh, and
  Mohamed]{SaufiSulaiman2019}
M.~{Saufi Sulaiman}, B.~Singh, and W.A.N.W. Mohamed.
\newblock Experimental and theoretical study of thermoelectric generator waste
  heat recovery model for an ultra-low temperature pem fuel cell powered
  vehicle.
\newblock \emph{Energy}, 179:\penalty0 628--646, 2019.
\newblock \doi{https://doi.org/10.1016/j.energy.2019.05.022}.

\bibitem[Wang et~al.(2016)Wang, Sanders, Dubey, Choo, and Duan]{Wang2016}
Kai Wang, Seth~R. Sanders, Swapnil Dubey, Fook~Hoong Choo, and Fei Duan.
\newblock Stirling cycle engines for recovering low and moderate temperature
  heat: A review.
\newblock \emph{Renewable and Sustainable Energy Reviews}, 62:\penalty0
  89--108, 2016.
\newblock \doi{10.1016/j.rser.2016.04.031}.

\bibitem[Cao et~al.(2023)Cao, Zheng, Zheng, Peng, Hu, Wang, and
  Leung]{Cao2023b}
Jingyu Cao, Ling Zheng, Zhanying Zheng, Jinqing Peng, Mingke Hu, Qiliang Wang,
  and Michael K.~H. Leung.
\newblock Recent progress in organic {Rankine} cycle targeting utilisation of
  ultra-low-temperature heat towards carbon neutrality.
\newblock \emph{Applied Thermal Engineering}, 231:\penalty0 120903, 2023.
\newblock \doi{10.1016/j.applthermaleng.2023.120903}.

\bibitem[Buehler et~al.(1963)Buehler, Gilfrich, and Wiley]{Buehler1963}
W.~J. Buehler, J.~V. Gilfrich, and R.~C. Wiley.
\newblock Effect of low-temperature phase changes on the mechanical properties
  of alloys near composition {TiNi}.
\newblock \emph{Journal of Applied Physics}, 34\penalty0 (5):\penalty0
  1475--1477, 1963.
\newblock \doi{10.1063/1.1729603}.

\bibitem[Chluba et~al.(2015)Chluba, Ge, de~Miranda, Strobel, Kienle, Quandt,
  and Wuttig]{Chluba2015}
Christoph Chluba, Wenwei Ge, Rodrigo~Lima de~Miranda, Julian Strobel, Lorenz
  Kienle, Eckhard Quandt, and Manfred Wuttig.
\newblock Ultralow-fatigue shape memory alloy films.
\newblock \emph{Science}, 348\penalty0 (6238):\penalty0 1004--1007, 2015.
\newblock \doi{10.1126/science.1261164}.

\bibitem[Hou et~al.(2019)Hou, Simsek, Ma, Johnson, Qian, Ciss{\'{e}}, Stasak,
  Hasan, Zhou, Hwang, Radermacher, Levitas, Kramer, Zaeem, Stebner, Ott, Cui,
  and Takeuchi]{Hou2019}
Huilong Hou, Emrah Simsek, Tao Ma, Nathan~S. Johnson, Suxin Qian, Cheikh
  Ciss{\'{e}}, Drew Stasak, Naila~Al Hasan, Lin Zhou, Yunho Hwang, Reinhard
  Radermacher, Valery~I. Levitas, Matthew~J. Kramer, Mohsen~Asle Zaeem,
  Aaron~P. Stebner, Ryan~T. Ott, Jun Cui, and Ichiro Takeuchi.
\newblock Fatigue-resistant high-performance elastocaloric materials made by
  additive manufacturing.
\newblock \emph{Science}, 366\penalty0 (6469):\penalty0 1116--1121, 2019.
\newblock \doi{10.1126/science.aax7616}.

\bibitem[Otsuka and Wayman(1999)]{Otsuka1999}
K.~Otsuka and C.~M. Wayman.
\newblock \emph{Shape Memory Materials}.
\newblock Cambridge University Press, Cambridge, 1999.
\newblock ISBN 9780521663847.

\bibitem[Huber et~al.(1997)Huber, Fleck, and Ashby]{Huber1997}
J.~E. Huber, N.~A. Fleck, and M.~F. Ashby.
\newblock The selection of mechanical actuators based on performance indices.
\newblock \emph{Proceedings of the Royal Society A: Mathematical, Physical and
  Engineering Sciences}, 453\penalty0 (1965):\penalty0 2185--2205, 1997.
\newblock \doi{10.1098/rspa.1997.0117}.

\bibitem[Johnson(1976)]{Johnson1976}
Alfred~Davis Johnson.
\newblock Memory alloy heat engine and method of operation.
\newblock U.S. Patent 4,055,955, 1976.

\bibitem[Sakuma and Iwata(1998)]{SAKUMA1998}
Toshio Sakuma and Uichi Iwata.
\newblock Working characteristics of a reciprocating-type heat engine using
  shape memory alloys.
\newblock \emph{JSME International Journal Series B}, 41\penalty0 (2):\penalty0
  344--350, 1998.
\newblock \doi{10.1299/jsmeb.41.344}.

\bibitem[Neumann and F{\"a}hler(2020)]{Neumann2020}
B.~Neumann and S.~F{\"a}hler.
\newblock Einrichtung zur nutzung thermischer energie.
\newblock Patent DE 10 2020 118 363.1 (filed 2020), 2020.

\bibitem[Neumann and F{\"a}hler(2024)]{Neumann2024}
B.~Neumann and S.~F{\"a}hler.
\newblock Vorrichtung zur wandlung thermischer fluidstr{\"o}mung und
  mechanischer arbeit.
\newblock Patent DE 10 2023 208 685.9 (granted 2024), 2024.

\bibitem[Neumann et~al.(2025)Neumann, Jocobi, Izadi, Henschke, and
  F{\"a}hler]{Neumann2025a}
Bruno Neumann, Giovanna Jocobi, Ali Izadi, Andreas Henschke, and Sebastian
  F{\"a}hler.
\newblock The power of thermoelastic harvesting of low-grade waste heat: A
  question of timing the heat exchange.
\newblock \emph{APL Energy}, 3\penalty0 (4), 2025.
\newblock \doi{10.1063/5.0293442}.

\bibitem[Neumann and F{\"a}hler(2025)]{Neumann2025}
Bruno Neumann and Sebastian F{\"a}hler.
\newblock Design guidelines for efficient thermoelastic harvesting of low-grade
  waste heat.
\newblock \emph{Energy Conversion and Management: X}, 27:\penalty0 101099,
  2025.
\newblock \doi{10.1016/j.ecmx.2025.101099}.

\bibitem[Chapman(2012)]{Chapman2012}
Stephen~J. Chapman.
\newblock \emph{Electric Machinery Fundamentals}.
\newblock McGraw-Hill, 5 edition, 2012.

\bibitem[Frenzel et~al.(2015)Frenzel, Wieczorek, Opahle, Maa{\ss}, Drautz, and
  Eggeler]{Frenzel2015}
Jan Frenzel, Andreas Wieczorek, Ingo Opahle, Bj{\"o}rn Maa{\ss}, Ralf Drautz,
  and Gunther Eggeler.
\newblock On the effect of alloy composition on martensite start temperatures
  and latent heats in {Ni}--{Ti}-based shape memory alloys.
\newblock \emph{Acta Materialia}, 90:\penalty0 213--231, 2015.
\newblock \doi{10.1016/j.actamat.2015.02.029}.

\bibitem[Khalil-Allafi et~al.(2002)Khalil-Allafi, Dlouhy, and
  Eggeler]{KhalilAllafi2002}
J.~Khalil-Allafi, A.~Dlouhy, and G.~Eggeler.
\newblock {Ni$_4$Ti$_3$}-precipitation during aging of {NiTi} shape memory
  alloys and its influence on martensitic phase transformations.
\newblock \emph{Acta Materialia}, 50\penalty0 (17):\penalty0 4255--4274, 2002.
\newblock \doi{10.1016/S1359-6454(02)00257-4}.

\bibitem[Liu et~al.(2023)Liu, Chen, Huang, Qiao, Yu, Xie, Ramanujan, Hu, Chu,
  Long, and Zhang]{Liu2023}
Xianliang Liu, Haodong Chen, Jianyi Huang, Kaiming Qiao, Ziyuan Yu, Longlong
  Xie, Raju~V. Ramanujan, Fengxia Hu, Ke~Chu, Yi~Long, and Hu~Zhang.
\newblock High-performance thermomagnetic generator controlled by a
  magnetocaloric switch.
\newblock \emph{Nature Communications}, 14\penalty0 (1):\penalty0 4811, 2023.
\newblock \doi{10.1038/s41467-023-40634-x}.

\bibitem[Snyder and Toberer(2008)]{Snyder2008}
G.~Jeffrey Snyder and Eric~S. Toberer.
\newblock Complex thermoelectric materials.
\newblock \emph{Nature Materials}, 7\penalty0 (2):\penalty0 105--114, 2008.
\newblock \doi{10.1038/nmat2090}.

\bibitem[Casi et~al.(2021)Casi, Araiz, Catal{\'a}n, and Astrain]{Casi2021}
{\'A}lvaro Casi, Miguel Araiz, Leyre Catal{\'a}n, and David Astrain.
\newblock Thermoelectric heat recovery in a real industry: From laboratory
  optimization to reality.
\newblock \emph{Applied Thermal Engineering}, 184:\penalty0 116275, 2021.
\newblock \doi{https://doi.org/10.1016/j.applthermaleng.2020.116275}.

\bibitem[{Shanghai Metals Market}(2026)]{metals}
{Shanghai Metals Market}.
\newblock \url{https://www.metal.com/}, 2026.
\newblock Accessed: 2026-01-07.

\bibitem[{Fraunhofer Institute for Solar Energy Systems
  ISE}(2026)]{FraunhoferISE2026}
{Fraunhofer Institute for Solar Energy Systems ISE}.
\newblock Photovoltaics report.
\newblock
  \url{https://www.ise.fraunhofer.de/content/dam/ise/de/documents/publications/studies/Photovoltaics-Report.pdf},
  2026.
\newblock Updated 22 June 2026; retrieved 2026-07-01.

\bibitem[Curzon and Ahlborn(1975)]{Curzon1975}
F.~L. Curzon and B.~Ahlborn.
\newblock Efficiency of a {Carnot} engine at maximum power output.
\newblock \emph{American Journal of Physics}, 43\penalty0 (1):\penalty0 22--24,
  1975.
\newblock \doi{10.1119/1.10023}.

\bibitem[Dzekan et~al.(2021)Dzekan, Waske, Nielsch, and F{\"a}hler]{Dzekan2021}
Daniel Dzekan, Anja Waske, Kornelius Nielsch, and Sebastian F{\"a}hler.
\newblock Efficient and affordable thermomagnetic materials for harvesting low
  grade waste heat.
\newblock \emph{{APL} Materials}, 9:\penalty0 011105, 2021.
\newblock \doi{10.1063/5.0033970}.

\bibitem[Frenzel et~al.(2018)Frenzel, Eggeler, Quandt, Seelecke, and
  Kohl]{Frenzel2018}
Jan Frenzel, Gunther Eggeler, Eckhard Quandt, Stefan Seelecke, and Manfred
  Kohl.
\newblock High-performance elastocaloric materials for the engineering of bulk-
  and micro-cooling devices.
\newblock \emph{MRS Bulletin}, 43\penalty0 (4):\penalty0 280--284, 2018.
\newblock \doi{10.1557/mrs.2018.67}.

\bibitem[Wieczorek et~al.(2017)Wieczorek, Frenzel, Schmidt, Maa{\ss}, Seelecke,
  Sch{\"u}tze, and Eggeler]{Wieczorek2017}
A.~Wieczorek, J.~Frenzel, M.~Schmidt, B.~Maa{\ss}, S.~Seelecke, A.~Sch{\"u}tze,
  and G.~Eggeler.
\newblock Optimizing {Ni}--{Ti}-based shape memory alloys for ferroic cooling.
\newblock \emph{Functional Materials Letters}, 10\penalty0 (01):\penalty0
  1740001, 2017.
\newblock \doi{10.1142/S179360471740001X}.

\bibitem[Lollobrigida et~al.(2014)Lollobrigida, Basso, Borgatti, Torelli,
  Kuepferling, Co\"isson, Olivetti, Celegato, Tortora, Stefani, Panaccione, and
  Offi]{Lollobrigida2014}
V.~Lollobrigida, V.~Basso, F.~Borgatti, P.~Torelli, M.~Kuepferling,
  M.~Co\"isson, E.~S. Olivetti, F.~Celegato, L.~Tortora, G.~Stefani,
  G.~Panaccione, and F.~Offi.
\newblock Chemical, electronic, and magnetic structure of {LaFeCoSi} alloy:
  Surface and bulk properties.
\newblock \emph{Journal of Applied Physics}, 115\penalty0 (20):\penalty0
  203901, 2014.
\newblock \doi{10.1063/1.4879195}.

\bibitem[Lim et~al.(2002)Lim, Xue, and Wang]{Lim2002}
J.~Lim, J.~M. Xue, and J.~Wang.
\newblock Ferroelectric lead scandium tantalate from mechanical activation of
  mixed oxides.
\newblock \emph{Materials Chemistry and Physics}, 75\penalty0 (1--3):\penalty0
  157--160, 2002.
\newblock \doi{10.1016/S0254-0584(02)00046-9}.

\end{thebibliography}

\end{document}